\documentclass[nofootinbib,twocolumn,prd,groupedaddress,preprintnumbers]{revtex4-1}
\usepackage{slashed}
\usepackage{bm}
\usepackage{graphicx}
\usepackage{color}
\usepackage{upgreek}
\usepackage{amssymb}
\usepackage{mathrsfs}
\usepackage{amsmath}
\usepackage[normalem]{ulem}
\usepackage{bbm}
\usepackage{array}
\RequirePackage[colorlinks=true,urlcolor=blue,anchorcolor=blue,citecolor=blue,filecolor=blue,
linkcolor=blue,menucolor=blue,linktocpage=true,pdfproducer=medialab]{hyperref}

\DeclareMathOperator{\diag}{diag}

\def\dim{\mathtt{d}}
\def\FF{\mathscr{F}}
\def\LL{\mathscr{L}}

\def\PP{\mathscr{P}}
\def\cD{{\cal D}}
\def\cF{{\cal F}}

\def\cO{{\cal O}}
\def\cP{{\cal P}}
\def\cY{{\cal Y}}
\def\U{\mathbf{U}}
\def\T{\mathbf{T}}
\def\V{\mathbf{V}}
\def\W{\mathbf{W}}
\def\D{\mathbf{D}}
\def\ct{\tilde{c}}

\newcommand{\DL}{\mathbf{D}}
\newcommand{\vh}{\langle \varphi \rangle}
\newcommand{\cAt}{\widetilde{{\cal A}}}

\newcommand{\Vt}{\widetilde{\mathbf{V}}}
\newcommand{\BLt}{\widetilde{\mathbf{B}}}
\newcommand{\WLt}{\widetilde{\mathbf{W}}}
\newcommand{\SH}{\mathbf{\Sigma}}

\newcommand{\WWd}{W_{\mu\nu}}

\newcommand{\BBd}{B_{\mu\nu}}
\newcommand{\BBu}{B^{\mu\nu}}
\newcommand{\GGd}{\mathcal{G}_{\mu\nu}}

\newcommand{\gtt}{\mathfrak{g}}
\newcommand{\htt}{\mathfrak{h}}

\def\nn{\nonumber}
\def\tr{{\rm Tr}}
\def\Tr{{\rm Tr}}
\def\hc{\text{h.c.}}
\def\derp{\partial}
\def\de{\partial}

\def\bc{\bar{\chi}}

\def\be{\begin{equation}}
\def\ee{\end{equation}}

\newcommand{\GeV}{\;\text{GeV}}

\newcommand{\blue}[1]{\color{blue} #1 \color{black} }

\begin{document}

\title{Distinguishing A Higgs-Like Dilaton Scenario With A Complete Bosonic Effective Field Theory Basis}%

\author{P. Hern\'andez-Le\'on}
\email{patricia.hernandezl@estudiante.uam.es}
\author{L. Merlo}
\email{luca.merlo@uam.es}
\affiliation{\vspace{1mm} 
Departamento de F\'isica Te\'orica and Instituto de F\'isica Te\'orica, IFT-UAM/CSIC, Universidad Aut\'onoma de Madrid, Cantoblanco, 28049, Madrid, Spain}

\begin{abstract}
A Higgs-like dilaton owns couplings that differ from those of the Standard Model Higgs and of a generic Composite Higgs. The complete bosonic basis for a Higgs-like dilaton is presented at the first subleading order. A comparison with the Standard Model, the Standard Model Effective Field Theory and the generic Lagrangian for the minimal $SO(5)/SO(4)$ Composite Higgs model is performed. Observables that can disentangle the different hypotheses are identified.

\end{abstract}

\date{\today}

\preprint{\blue{FTUAM-17-4}}
\preprint{\blue{IFT-UAM/CSIC-17-017}}

\maketitle

%
%
\section{Introduction}

The observation of a neutral, CP-even scalar particle at LHC~\cite{Aad:2012tfa,Chatrchyan:2012xdj} represents a unique opportunity to investigate the scalar sector of the Standard Model (SM) and to shed light on the electroweak symmetry breaking  (EWSB) mechanism. At present, there is no evidence for deviations from the SM scalar boson (``Higgs'' for short) hypothesis -- see Ref.~\cite{Mariotti:2016owy} for a recent update. However, other theories with Higgs-like candidates are still equally viable within the present sensitivities: models with more than one Higgs doublet, such as supersymmetry or two-Higgs-doublet constructions; Composite Higgs (CH) theories; models where the Higgs arises as a Goldstone boson. Disentangling the different alternatives, considering the various experimental facilities, is crucial.

This has been deeply pursued adopting the Effective Field Theory approach~\cite{Weinberg:1978kz}, which allows to avoid the specificity of the distinct models and, instead, provides model-independent predictions characteristic of more general frameworks. Models where the physical Higgs particle belongs to a doublet representation of the electroweak (EW) symmetry can be described at low-energy by the so-called Standard Model Effective Field Theory (SMEFT) Lagrangian~\cite{Buchmuller:1985jz,Grzadkowski:2010es}. Otherwise, a more suitable description is the so-called Higgs Effective Field Theory Lagrangian (HEFT) Lagrangian~\cite{Feruglio:1992wf,Contino:2010mh,Alonso:2012px,Alonso:2012pz,Buchalla:2013rka,Gavela:2014vra}. 

The two effective Lagrangians provide a description of gauge, fermion and Higgs couplings, respecting Lorentz and $SU(3)_c\times SU(2)_L\times U(1)_Y$ gauge invariance. The key difference between the two approaches is the relationship between the physical Higgs field $h(x)$ and the SM Goldstone bosons (GBs) $\overrightarrow{\pi}(x)$: in the SMEFT, the four fields belong to the $SU(2)_L$ doublet $\Phi(x)$,
\begin{equation}
\Phi(x)=\U(x) \begin{pmatrix} 0 \\ \frac{v+h(x)}{\sqrt{2}} \end{pmatrix}\,,
\label{HiggsEq}
\end{equation}
with
\be
\U(x)\equiv e^{i\overrightarrow\sigma\cdot\overrightarrow\pi(x)/v}
\ee
being the GB matrix and $v$ the EW vacuum expectation value (VEV) fixed through the $W$ mass. On the contrary, in the HEFT the physical Higgs and the GB matrix are treated as independent objects~\cite{Feruglio:1992wf,Contino:2010mh,Alonso:2012jc,Alonso:2012px,Alonso:2012pz,Brivio:2013pma,Buchalla:2013rka,Brivio:2014pfa,Gavela:2014vra,Brivio:2015kia,Brivio:2016fzo,Gavela:2014uta,Merlo:2016prs,Brivio:2017ije}, which leads to a much larger number of operators in the HEFT with respect to the SMEFT, at the same order in the expansion~\cite{Manohar:1983md,Cohen:1997rt,Gavela:2016bzc}. Moreover, from the dimensionlessness of the GB matrix it follows a reshuffling of the leading operators in the HEFT with respect to the SMEFT Lagrangian. In Refs.~\cite{Brivio:2013pma,Brivio:2014pfa,Brivio:2015kia,Brivio:2016fzo,Eboli:2016kko} it has been shown that HEFT exhibits the following specific features:
\begin{itemize}
\item[-] some correlations typical of the SMEFT, such as those between triple and quartic gauge couplings, are lost in the HEFT;
\item[-] Higgs couplings to gauge bosons are correlated to pure gauge couplings in the SMEFT, while they are completely free in the HEFT;
\item[-] a few couplings that are expected to be strongly suppressed in the SMEFT, are instead predicted with higher strength in the HEFT and potentially lead to visible observables in the present LHC run.
\end{itemize}

The HEFT Lagrangian is a very useful tool to describe an extended class of ``Higgs'' models: by fixing the Lagrangian parameters, it can encode SM and SMEFT, Goldstone Boson Higgs models~\cite{Kaplan:1983fs,Kaplan:1983sm,Banks:1984gj,Agashe:2004rs,Gripaios:2009pe,Feruglio:2016zvt,Gavela:2016vte} and dilaton-like constructions~\cite{Halyo:1991pc,Goldberger:2008zz,Vecchi:2010gj,Matsuzaki:2012mk,Chacko:2012sy,Chacko:2012vm,Bellazzini:2012vz}. Therefore, HEFT can be considered the most general description of gauge, fermion and Higgs couplings, invariant under $SU(3)_c\times SU(2)_L\times U(1)_Y$ gauge symmetry.

The comparison between HEFT and SMEFT Lagrangians has undergone an intensive investigation~\cite{Brivio:2013pma,Brivio:2014pfa,Brivio:2015kia,Brivio:2016fzo,Eboli:2016kko}. As well, the matching between CH models and the HEFT also received much attention: in particular, in Refs.~\cite{Alonso:2014wta,Hierro:2015nna}, considering a CH model with a symmetric coset, the potentiality of the HEFT Lagrangian has been shown to account for Composite Higgs models as a possible ultraviolet completion. 

On the other hand, the link with dilaton constructions has been less studied, even if the dilaton solution to the Hierarchy problem is attractive and largely investigated~\cite{Halyo:1991pc,Goldberger:2008zz,Vecchi:2010gj,Matsuzaki:2012mk,Chacko:2012vm,Chacko:2012sy,Bellazzini:2012vz}. Moreover, recent lattice simulations of strongly interacting gauge theories predict the appearance of a scalar particle that could be interpreted as a dilaton, when the conformal behaviour sets in~\cite{Fodor:2012ty,Aoki:2014oha,Appelquist:2016viq,Fodor:2016pls,Aoki:2016wnc,Appelquist:2017wcg}. Dilaton models are based on the fact that the SM Lagrangian is approximately scale invariant, once neglecting explicit scales associated to the EWSB mechanism and the dynamical conformal breaking due to QCD. In models where the scale invariance is broken spontaneously, a GB naturally arises~\cite{Hellerman:2015nra,Monin:2016jmo} and can then be identified with the physical Higgs field: its mass is then protected by the GB shift symmetry and can only acquire (relatively small) values, close to the symmetry breaking scale.

It is the aim of the present paper to complete the comparison between the distinct Higgs setups: SM, SMEFT, CH models and dilaton theories. For definiteness, the minimal $SO(5)/SO(4)$ CH model will be considered in the following, even if the results hold also for other models, such as for the original $SU(5)/SO(5)$ model by Georgi and Kaplan~\cite{Kaplan:1983fs}.  As couplings with fermions follow specific assumptions on the underlying framework, the analysis will focus on physical effects in the bosonic sector only. Moreover, the analysis will consider only CP-even couplings, as they suffice to illustrate the general features of the comparison between the distinct setups.

The structure of the paper is as follows. First, the HEFT framework is summarised in Sect.~\ref{Sect:HEFT}. Then, the most generic effective Lagrangian for a Higgs-like dilaton is constructed in Sect.~\ref{Sect:DilatonBasis}. In Sect.~\ref{Sect:Comparison}, the comparison with SM, SMEFT and the minimal $SO(5)/SO(4)$ CH model is presented, including the discussion on possible discriminating signals. Concluding remarks are provided in Sect.~\ref{Sect:Conclusions}. App.~\ref{APP:Counting} contains the indications of the constructions of the HEFT and Dilaton Effective Field Theory (DEFT) Lagrangians. In Apps.~\ref{APP:SMEFT} and \ref{APP:CH}, the SMEFT Lagrangian and the Lagrangian for the minimal $SO(5)/SO(4)$ CH model, respectively, are briefly summarised. 

%
%
\section{The HEFT Lagrangian}
\label{Sect:HEFT}

The building blocks used to construct the HEFT are the SM fermions and gauge bosons, together with the GB matrix $\U(x)$ and the physical Higgs particle $h(x)$. The GB matrix transforms as a bi-doublet under the global $SU(2)_L\times SU(2)_R$ symmetry,
\be
\U(x)\rightarrow L\, \U(x)\, R^\dag\,,
\label{Umatrix}
\ee
where $L$ and $R$ are the unitary transformations of $SU(2)_L$ and $SU(2)_R$, respectively. It is typically encoded into two chiral fields
\be
\begin{aligned}
&\T(x) \equiv \U(x) \sigma_3 \U^\dagger(x)\,,\quad 
&&\T(x)\rightarrow L\,\T(x)L^\dagger\,,\\
&\V_\mu(x) \equiv \left(\D_\mu \U(x)\right)\U^\dagger(x)\,,\quad
&&\V_\mu(x)\rightarrow L\,\V_\mu(x)L^\dagger\,,
\end{aligned}
\ee
with 
\be
\D_\mu \U(x) \equiv \derp_\mu \U(x) +ig\W_{\mu}(x)\U(x) - \dfrac{ig'}{2} B_\mu(x) \U(x)\sigma_3 \,,
\ee
where $\W_\mu(x)\equiv W_{\mu}^a(x)\sigma_a/2$. While both fields transform in the adjoint of $SU(2)_L$, the scalar field $\T(x)$ breaks explicitly the $SU(2)_R$ symmetry and thus, when present in an operator, allows an easy identification of its $SU(2)_R$ non-conserving nature. The Lagrangian is approximatively written invariant under the global $SU(2)_L\times SU(2)_R$ symmetry: as in the SM, it is spontaneously broken down to the $SU(2)_C$ custodial symmetry and explicitly broken by the gauging of Hypercharge and the heterogeneity of the fermion masses; some operators containing $\T$ are indeed associated to the Hypercharge; others represent instead sources of custodial breaking beyond the SM ones. 

The physical Higgs $h(x)$ is an isosinglet of the SM gauge symmetry and is conventionally described through dimensionless  generic functions $\cF(h/v)$~\cite{Feruglio:1992wf,Grinstein:2007iv}, with $v\approx246\GeV$ the EW scale. Distinct functions $\cF(h)$ identify scalar field manifolds with different curvatures~\cite{Alonso:2015fsp,Alonso:2016btr,Alonso:2016oah}, which represent an observable measurable at LHC. 

The $\cF(h/v)$ functions are commonly written as a polynomial expansion in $h/v$, $\cF(h/v)=1+\alpha (h/v)+\beta (h/v)^2+\ldots$, where dots account for higher powers of $(h/v)$. In specific realisations, the scale associated to $h$ can be distinct from $v$: in Composite Higgs models, for example, this scale is larger than $v$ and is associated to the scale at which the GBs arise after the spontaneous breaking of the initial global symmetry. According to the Naive Dimensional Analysis (NDA)~\cite{Manohar:1983md,Cohen:1997rt,Gavela:2016bzc}, which determines the suppressions of the distinct effective operators, the scale associated to $h$, usually denoted by $f$, satisfies $f\neq v$ and $ \Lambda\leq4\pi f$, being $\Lambda$ the scale that fixes the validity of the theory. However, in the HEFT formalism~\cite{Gavela:2016bzc,Brivio:2016fzo}, factors of $v/f$, being $f$ this new scale, are accounted for in the free coefficients of the effective Lagrangian, \mbox{$\alpha$, $\beta$, \ldots,} leaving $v$ as the only sensitive scale in the $\cF(h/v)$ functions. In what follows the notation will be simplified, suppressing the explicit dependence on $v$. 

Finally, SM fermions are arranged in doublets of the global $SU(2)_L$ or $SU(2)_R$ symmetries: in particular, the right-handed fields are collected in the following spinors,
\be
Q_{R}=\begin{pmatrix} u_{R} \\ d_{R} \end{pmatrix}\,, \qquad\qquad 
L_{R}=\begin{pmatrix} N_{R} \\ e_{R} \end{pmatrix}\,,
\label{doublets}
\ee
where $N_{R}$ are three right-handed neutrinos, introduced here to complete the $SU(2)_R$ doublet\footnote{The discussion on neutrino masses will not be treated here. See Ref.~\cite{Hirn:2005fr,Merlo:2016prs} for details.}. 

Following Ref.~\cite{Brivio:2016fzo} and the discussion in App.~\ref{APP:Counting}, the HEFT Lagrangian can be written as a sum of two terms,
\be
\LL_\text{HEFT}\equiv \LL_{h}^{(0)} + \Delta \LL_{h}\,,
\label{HEFTLAG}
\ee
where the first one reads:
\begin{align}
\LL_h^{(0)}=& -\dfrac{1}{4} \BBd\BBu -\dfrac{1}{4} \WWd^a W^{a\,\mu\nu}-\dfrac{1}{4} \GGd^\alpha \mathcal{G}^{\alpha\,\mu\nu}+\nn\\
&+\dfrac{1}{2}\de_\mu h \de^\mu h-\dfrac{v^2}{4}\Tr(\V_\mu \V^\mu)\cF_C(h)-V(h)
+\nn\\
&+i\bar{Q}_L\slashed{D}Q_L+i\bar{Q}_R\slashed{D}Q_R+i\bar{L}_L\slashed{D}L_L
+i\bar{L}_R\slashed{D}L_R+\nn\\
&-\dfrac{v}{\sqrt2}\left(\bar{Q}_L\U \cY_Q(h) Q_R+\hc\right)+\label{Lag0}\\
&-\dfrac{v}{\sqrt2}\left(\bar{L}_L\U \cY_L(h) L_R+\hc\right)+\nn\\
&-\dfrac{g_s^2}{(4\pi)^2}\lambda_s\,\GGd^\alpha\,
\tilde{\mathcal{G}}^{\alpha\,\mu\nu}\,,\nn
\end{align}
where $\tilde{\mathcal{G}}^{\mu\nu}\equiv \frac12\epsilon^{\mu\nu\rho\sigma} \mathcal{G}_{\rho\sigma}$. It contains the kinetic terms for all the fields, the theta term of QCD, the mass terms for the EW gauge bosons, the Yukawa interactions and the Higgs scalar potential, whose specific form depends on the model under consideration~\cite{Bellazzini:2014yua}. 
Notice that the operator $\Tr(\V_\mu \V^\mu)\cF_C(h)$ is multiplied by $v^2$, while according to NDA it should be multiplied by $f^2$. Similarly, the Yukawa-like interactions are multiplied by $v$ instead of by $f$. This is the well-known fine-tuning of the HEFT. However, this should not be taken as a reason to consider the HEFT unable to describe the EW and Higgs interactions; instead, this should be interpreted as an indication that the underlying theory which projects into the HEFT at low-energies should account for a mechanism to predict the correct scale for the EW gauge boson and fermion masses. An example is provided by CH models, where the SM gauge boson masses are predicted at the EW scale as a natural feature of these theories. Fermion masses are described at the EW scale once considering the so-called fermion partial compositeness mechanism. 

The functions $\cY_{Q,L}(h)$ appearing in the Yukawa couplings are written in a compact notation and with the flavour indices left implicit:
\begin{equation}
\begin{aligned}
\cY_{Q}(h)\equiv& \diag\left(\sum_n Y_{U}^{(n)}\dfrac{h^n}{v^n},
\sum_n Y_{D}^{(n)}\dfrac{h^n}{v^n}\right)\,,\\
\cY_{L}(h)\equiv& \diag\left(0,\sum_n Y_{\ell}^{(n)}\dfrac{h^n}{v^n}\right)\,.
\end{aligned}
\end{equation}
The $n=0$ terms yield fermion masses, while the higher orders describe the interactions with $n$ insertions of the Higgs field $h$, accounting in general for non-aligned contributions. The structure of these terms, however, is customarily simplified in phenomenological analyses, assuming Yukawa interactions aligned with fermion masses, i.e. $Y_{U,D,\ell}^{(n)}=Y_{U,D,\ell}^{(0)}$~\cite{Brivio:2013pma,Gavela:2014vra,Brivio:2015kia,Brivio:2016fzo}. The same assumption will be adopted here and in consequence 
\be
\begin{aligned}
\cY_{Q}(h)&=\diag\left(Y_{U}^{(0)}\cF_U(h),Y_{D}^{(0)}\cF_D(h)\right)\,,\\
\cY_{L}(h)&=\diag\left(0,Y_{\ell}^{(0)}\cF_\ell(h)\right)\,.
\end{aligned}
\ee
Notice that, the heterogeneity of fermion masses encoded into $\cY_{Q}(h)$ and $\cY_{L}(h)$ breaks explicitly the custodial symmetry.

The function $\cF_C(h)$ appearing in the GB kinetic term is typically expanded as
\begin{equation}
\cF_C(h) = 1 + 2 a_C \frac{h}{v}+b_C \frac{h^2}{v^2}+\dots
\label{FC}
\end{equation}
It is convenient to make explicit the beyond SM part of the coefficients $a_C,\, b_C$, using the notation
\begin{equation}
 a_C = 1 + \Delta a_C\,,\qquad 
 b_C = 1 + \Delta b_C\,,
\end{equation} 
where $\Delta a_C,\, \Delta b_C$ are assumed to be of the same order as the coefficients accompanying the operators appearing in the second part of the Lagrangian $\Delta \LL_h$, which accounts for new interactions and for deviations from the leading order (LO) one. 

$\cF(h)$ functions could be inserted into the kinetic terms for fermions and the physical Higgs, but their contributions can be reabsorbed inside the generic functions $\cF_C(h)$ and $\cY_{Q,L}(h)$, as discussed in Ref.~\cite{Brivio:2016fzo}. The kinetic terms of the gauge bosons are also free from any $\cF(h)$ in this LO Lagrangian, assuming that the transverse components of the gauge bosons do not couple strongly to the EWSB sector~\cite{Contino:2010mh}.

Following Refs.~\cite{Brivio:2013pma,Brivio:2016fzo} and the discussion in App.~\ref{APP:Counting}, all the operators necessary for reabsorbing 1-loop divergences arising from the renormalisation of $\LL_h^{(0)}$ are contained in the second part of the Lagrangian in Eq.~(\ref{HEFTLAG}). Adopting the notation used in Refs.~\cite{Alonso:2014wta,Hierro:2015nna} and the NDA normalisation~\cite{Manohar:1983md,Cohen:1997rt,Gavela:2016bzc}, $\Delta\LL_h$ can be written as the sum of several terms: focussing only on the CP-even bosonic ones, one can write
\begin{align}
\Delta\LL_h= & c_T \cP_T\cF_T(h)+c_B\cP_B \cF_B(h)+\nn\\
&+ c_W\cP_W\cF_W(h)+c_G\cP_G\cF_G(h) +\nn\\
&+c_H \cP_H\cF_H(h)+c_{\Box H} \cP_{\Box H}\cF_{\Box H}(h)+\nn\\
&+c_{\Delta H} \cP_{\Delta H}\cF_{\Delta H}(h)+c_{DH} \cP_{DH}\cF_{DH}(h)+\nn\\
&+c_{WWW}\cP_{WWW}\cF_{WWW}(h)+\label{DeltaL}\\
&+c_{GGG}\cP_{GGG}\cF_{GGG}(h)+c_{DB} \cP_{DB}\cF_{DB}(h)+\nn\\
&+c_{DW} \cP_{DW}\cF_{DW}(h)+c_{DG} \cP_{DG}\cF_{DG}(h)+\nn\\
&+\sum_{i=1}^{26} c_i\cP_i\cF_i(h)\,,\nn
\end{align}
where the parameters $c_i$ are free coefficients smaller than 1, according to the Naive Dimensional Analysis formulation~\cite{Manohar:1983md,Cohen:1997rt,Gavela:2016bzc}.
The term $\cP_T$ is a custodial-breaking two-derivative operator,
\be
\cP_T = \frac{v^2}{4} \tr(\T\V_\mu)\tr(\T\V^\mu) \,,
\label{P_T}
\ee
that traditionally is inserted at the LO, but whose coefficient is so strongly constrained ($\lesssim 10^{-2}$) from the bounds on the EW precision parameter $T$ that it is customarily listed in $\Delta\LL_h$. The three operators $\cP_B$, $\cP_W$ and $\cP_G$,
\be
\begin{aligned}
\cP_{B} &=-\frac{1}{4}\BBd \BBu \\
\cP_{W} &=-\frac{1}{4}\WWd^a W^{a\,\mu\nu} \\
\cP_G &= -\frac{1}{4}\GGd^\alpha\, \mathcal{G}^{\alpha\,\mu\nu}\,,
\end{aligned}
\ee
contain the field strengths for the SM gauge bosons and, once multiplied by the corresponding $\cF(h)$, describe the interactions between $h$ and the transverse componentes of the gauge bosons.

The four operators $\cP_{H}$, $\cP_{\Box H}$, $\cP_{\Delta H}$ and $\cP_{D H}$,
\be
\begin{aligned}
\cP_H &= \frac{1}{2}(\derp_\mu h)(\derp^\mu h)\\
\cP_{\Box H}&=\frac{1}{\Lambda^2}(\Box h)^2\\
\cP_{\Delta H}&=\frac{4\pi}{\Lambda^3}(\derp_\mu h)(\derp^\mu h)(\Box h)\\
\cP_{DH}&=\frac{(4\pi)^2}{\Lambda^4} \left((\derp_\mu h)(\derp^\mu h)\right)^2\,,
\end{aligned}
\label{SelfHiggsOperators}
\ee
describe pure Higgs couplings with two and four derivatives. 

The five pure-gauge operators $\cP_{WWW}$, $\cP_{GGG}$, $\cP_{DB}$, $\cP_{DW}$ and $\cP_{DG}$,
\be
\begin{aligned}
\cP_{WWW} &=\dfrac{4\pi\varepsilon_{abc}}{\Lambda^2}W_\mu^{a\nu}W_\nu^{b\rho} W_{\rho}^{c\mu}\,,\\
\cP_{GGG} &=\dfrac{4\pi f_{\alpha\beta\gamma}}{\Lambda^2}{\mathcal G}_\mu^{\alpha\nu}{\mathcal G}_\nu^{\beta\rho} {\mathcal G}_{\rho}^{\gamma\mu}\\
\cP_{DB} & = \dfrac{1}{\Lambda^2}\left(\derp^\mu B_{\mu\nu}\right)\left(\derp_\rho B^{\rho\nu}\right)\\
\cP_{DW} & = \dfrac{1}{\Lambda^2}\left(\cD^\mu W_{\mu\nu}\right)^a\left(\cD_\rho W^{\rho\nu}\right)^a\\
\cP_{DG} & = \dfrac{1}{\Lambda^2}\left(\cD^\mu G_{\mu\nu}\right)^\alpha\left(\cD_\rho G^{\rho\nu}\right)^\alpha\,,
\end{aligned}
\ee
with $\varepsilon_{abc}$ and $f_{\alpha\beta\gamma}$ the structure constants of $SU(2)_L$ and $SU(3)_c$ respectively, are typically listed at higher orders in the chiral perturbation theory, but in the HEFT they should be inserted in $\Delta\LL_h$: indeed, they have the same suppressions of $\cP_{\square H}$ in Eq.~(\ref{SelfHiggsOperators}) (and of four-fermion operators, if fermion couplings would be considered -- see Ref.~\cite{Brivio:2016fzo}). This is consistent with the fact that the HEFT merges together the traditional expansion in canonical dimensions and the expansion in derivatives of the chiral perturbation theory (see Ref.~\cite{Gavela:2016bzc} for details). 

The rest of the terms are four-derivative operators defined as
\begin{widetext}
\be
\begin{aligned}
\begin{aligned}
\cP_{1}  &=B_{\mu\nu} \Tr\left(\T\,\W^{\mu\nu}\right)\\
\cP_{2}  &=\dfrac{i}{4\pi}B_{\mu\nu} \Tr\left(\T\left[\V^\mu,\V^\nu\right]\right) \\
\cP_{3}  &= \dfrac{i}{4\pi}\Tr\left(\W_{\mu\nu}\left[\V^\mu,\V^\nu\right]\right)\\
\cP_{4}  &= \dfrac{i}{\Lambda}B_{\mu\nu}\Tr(\T\V^\mu)\,\derp^\nu h\\
\cP_{5}  &= \dfrac{i}{\Lambda}\Tr(\W_{\mu\nu}\V^\mu)\,\derp^\nu h \\
\cP_{6} &=\dfrac{1}{(4\pi)^2}\left(\Tr\left(\V_\mu\,\V^\mu\right)\right)^2 \\
\cP_{7} &=\dfrac{1}{4\pi\Lambda}\Tr\left(\V_\mu\,\V^\mu\right)\derp_\nu\derp^\nu h\\
\cP_{8} &=\dfrac{1}{\Lambda^2}\Tr\left(\V_\mu\,\V_\nu\right)\derp^\mu h\,\derp^\nu h\\
\cP_{9} &=\dfrac{1}{(4\pi)^2}\Tr\left((\cD_\mu\V^\mu)^2 \right)\\
\cP_{10} &=\dfrac{1}{4\pi\Lambda}\Tr(\V_\nu \,\cD_\mu\V^\mu)\,\derp^\nu h\\
\cP_{11} &=\dfrac{1}{(4\pi)^2}\left(\Tr\left(\V_\mu\,\V_\nu\right)\right)^2\\
\cP_{12} &=\left(\Tr\left(\T \W_{\mu\nu}\right)\right)^2\\
\cP_{13} &=\dfrac{i}{4\pi}\Tr\left(\T \W_{\mu\nu}\right)\Tr\left(\T\left[\V^\mu,\V^\nu\right]\right)
\end{aligned}\qquad\qquad
\begin{aligned}
\cP_{14} &=\dfrac{\varepsilon_{\mu\nu\rho\lambda}}{4\pi}\Tr\left(\T \V^\mu\right)\Tr\left(\V^\nu \W^{\rho\lambda}\right)\\
\cP_{15} &= \dfrac{1}{(4\pi)^2}\Tr(\T\,\cD_\mu\V^\mu)\,\Tr(\T\,\cD_\nu\V^\nu)\\
\cP_{16} & = \dfrac{1}{(4\pi)^2}\Tr([\T \,,\V_\nu]\,\cD_\mu \V^\mu) \, \Tr(\T\V^\nu) \\
\cP_{17} &= \dfrac{i}{\Lambda}\Tr(\T \W_{\mu\nu})\Tr(\T\V^\mu)\,\derp^\nu h\\
\cP_{18} &=\dfrac{1}{4\pi\Lambda}\Tr(\T\,[\V_\mu,\V_\nu])\Tr(\T\V^\mu)\,\derp^\nu h\\
\cP_{19} &=\dfrac{1}{4\pi\Lambda}\Tr(\T\,\cD_\mu\V^\mu)\Tr(\T\V_\nu)\,\derp^\nu h\\
\cP_{20} &=\dfrac{1}{\Lambda^2}\Tr\left(\V_\mu\,\V^\mu\right)\derp_\nu h\, \derp^\nu h\\
\cP_{21} &=\dfrac{1}{\Lambda^2}\left(\Tr\left(\T\,\V_\mu\right) \right)^2 \derp_\nu h\, \derp^\nu h\\
\cP_{22} &=\dfrac{1}{\Lambda^2}\Tr\left(\T\V_\mu\right)\Tr\left(\T\V_\nu\right)\,\derp^\mu h\,\derp^\nu h\\
\cP_{23} &=\dfrac{1}{(4\pi)^2} \Tr\left(\V_\mu\V^\mu\right)\left(\Tr\left(\T\V_\nu\right)\right)^2\\
\cP_{24} &=\dfrac{1}{(4\pi)^2}\Tr\left(\V_\mu\V_\nu\right)\Tr\left(\T\V^\mu\right)\Tr\left(\T\V^\nu\right) \\
\cP_{25} &=\dfrac{1}{4\pi\Lambda}\left(\Tr\left(\T\,\V_\mu\right) \right)^2 \derp_\nu\derp^\nu h\\
\cP_{26} &=\dfrac{1}{(4\pi)^2}\left(\Tr\left(\T\V_\mu\right)\Tr\left(\T\V_\nu\right)\right)^2\,,
\end{aligned}
\end{aligned}
\label{HEFTOperators}
\ee
\end{widetext}
where the normalisation follows the NDA of Refs.~\cite{Gavela:2016bzc}.

The operators of this list containing the scalar chiral field $\T$, not in association with the gauge field strength $B_{\mu\nu}$, represent sources of custodial symmetry breaking beyond those of the SM: they are $\cP_{12}-\cP_{19}$, and $\cP_{21}-\cP_{26}$. If one considers the custodial symmetry as a fundamental symmetry of the UV theory that gives rise to the HEFT Lagrangian at low-energy, this set of operators plus $\cP_T$ should be further suppressed. However, without assuming any specific underlying model, this is not justified, as shown for example in Ref.~\cite{Alonso:2014wta,Hierro:2015nna}. Here and in the following sections, the discussion will be kept general, without introducing any arbitrary suppression on these operators.\\

Once specifying an underlying scenario, it is then possible to write the Wilson coefficients $c_i$ and the functions $\cF_i(h)$ of the HEFT in terms of the parameters of the high-energy Lagrangian. In Refs.~\cite{Alonso:2014wta,Hierro:2015nna}, a generic Composite Higgs model has been considered and the corresponding Lagrangian at high-energy has been projected at low energy on the HEFT. The results can be read in Tables 1 and 2 of Ref.~\cite{Alonso:2014wta} and Table 1 of Ref.~\cite{Hierro:2015nna}. This exercise has pointed out that, in traditional Composite Higgs models, gauge-Higgs couplings are correlated to pure gauge ones, similarly to what happens in the SMEFT, although with a few differences: this follows from the fact that the Higgs field belongs to a representation that contains the $SU(2)_L$-doublet one~\cite{Panico:2015jxa}. The SMEFT is then a good low-energy description of Composite Higgs models at the very first order, but deviations arise at higher orders. As a result, one could disentangle an elementary from a composite Higgs comparing the same gauge interactions but with different Higgs legs. 

The dilaton, that will be the subject of the next section, is a singlet under $SU(2)_L$ and therefore distinct phenomenological features are expected: indeed, the EW doublet or singlet representation of the Higgs field is encoded into correlations/decorrelations between pure-gauge and gauge-Higgs couplings.

%
%
\section{The Higgs-Like Dilaton }
\label{Sect:DilatonBasis}

The dilaton arises as a GB of the spontaneous breaking of the scale invariance. This mechanism is typical of models where the EWSB and the scale symmetry breaking do not coincide, such as when the EWSB is strongly coupled: the scale symmetry breaking scale $f$ is typically larger than the EW VEV $v$ and the states arising from the scale symmetry breaking do not need to coincide with those responsible for the EWSB mechanism. Explicit realisations are theories of walking technicolor~\cite{Holdom:1984sk,Yamawaki:1985zg,Appelquist:1986an}, and of Randall-Sundrum extra-dimension constructions~\cite{Randall:1999ee,Contino:2003ve,Gherghetta:2003he,Agashe:2003zs}. Beside its GB nature, the dilaton can remain relatively light, although an explicit realisation is not easy to achieve~\cite{Chacko:2013dra,Bellazzini:2013fga,Coradeschi:2013gda,Megias:2014iwa}.

Considering a generic Lagrangian written as the sum of distinct operators $\cO_i(x)$ with canonical dimension $[\cO_i]\equiv \dim_i$ and with coefficient $g_i(\mu)$, where $\mu$ is the reference scale,
\begin{equation}
\LL = \sum_i g_i (\mu) {\cal O}_i(x)\,,
\label{eq:L}
\end{equation}
an infinitesimal scale transformation $x^\mu\to e^{\lambda} x^\mu$ gives
\be
\begin{aligned}
\cO_i(x) &\to e^{\lambda \dim_i} {\cal O}_i( e^{\lambda} x),\\
\mu &\to e^{-\lambda} \mu,
\end{aligned}
\ee
that leads to the variation of the Lagrangian
\begin{equation}
\label{eq:dL}
\delta \LL = \sum_i  g_i(\mu) (\dim_i + x^\mu\partial_\mu) \cO_i(x) + \sum_i {\beta_i}(g) \cO_i(x),
\end{equation}
where
\be
\beta_i(g)=-\mu  \dfrac{\derp g_i(\mu)}{\derp\mu}\,,
\ee
are the beta functions of the couplings $g_i$. The Lagrangian is scale invariant when $\delta \LL =0$, which corresponds to  ${\beta_i}(g)=0$, i.e. the couplings do not depend on the scale considered, and to the canonical dimensions of the operators satisfying to $\dim_i=4$ (by using integration by parts, $(\dim_i + x^\mu\partial_\mu) \cO_i(x)\to(\dim_i -4) \cO_i(x)$). See Ref.~\cite{ColemanBook} for a pedagogical introduction on the dilaton.

In the SM, fermion ($\dim=3$) and scalar ($\dim=2$) mass terms explicitly violate the scale symmetry. However, as adopted in the context of the so-called Minimal Flavour Violation~\cite{D'Ambrosio:2002ex,Cirigliano:2005ck,Davidson:2006bd,Grinstein:2010ve,Feldmann:2010yp,Alonso:2011yg,Guadagnoli:2011id,Alonso:2011jd,Buras:2011zb,Buras:2011wi,Alonso:2012fy,Lopez-Honorez:2013wla,Alonso:2013mca,Alonso:2016onw,Dinh:2017smk}, or more in general in flavour models~\cite{Froggatt:1978nt,Ma:2004zv,Altarelli:2005yp,Altarelli:2005yx,Feruglio:2007uu,Feruglio:2008ht,Bazzocchi:2009pv,Altarelli:2009gn,Altarelli:2010gt,Varzielas:2010mp,Toorop:2010ex,Altarelli:2012bn,Altarelli:2012ss,Bazzocchi:2012st,Altarelli:2012ia,Bergstrom:2014owa}, a strategy to reestablish scale invariance consists in enlarging the spectrum by the addition of a scalar field, $\chi(x)$, that transforms under the scale symmetry: insertion of powers of this scalar field in the Lagrangian operators then allows to recover exactly scale invariance. Once it develops a VEV, fermion and scalar masses are correctly described and the scale symmetry is spontaneously broken with the arising of the corresponding GB, the dilaton. 

Defining the scale transformation law for the additional scalar field $\chi(x)$ -- sometimes also called conformal compensator -- as
\be
\chi(x) \to e^{\lambda} \chi(e^{\lambda} x)\,,
\ee
scale symmetry invariance is recovered performing the following replacement for the couplings $g_i$ in Eq.~(\ref{eq:L}):
\be
g_i(\mu)\rightarrow g_i\left(\mu {\chi\over f}\right) \left({\chi\over f}\right)^{4-\dim_i}\,.
\label{ricetta}
\ee
The scale $f$ in the previous equation is identified with the VEV of $\chi(x)$, $f=\langle\chi\rangle$, and represents the scale of the symmetry breaking. Following the traditional notation for Goldstone bosons, the conformal compensator $\chi(x)$ can be parametrised as 
\be
\chi(x) = fe^{\sigma(x)/f}\,,
\label{sigmadef}
\ee
making explicit the GB $\sigma$, whose scale transformation is non-linear, $\sigma(x)/f\rightarrow \sigma(e^{\lambda} x)/f+\lambda$. 

To simplify the notation, the explicit dependence of $\chi(x)$ and $\sigma(x)$ on $x$ will be suppressed in what follows.

%
%
\subsection{The DEFT Lagrangian}
\label{Sect:DEFT}

The Dilaton Effective Field Theory (DEFT) Lagrangian is constructed in a very similar way to the HEFT one, by exchanging the Higgs field $h$ with the conformal compensator $\chi$, or the dilaton field $\sigma$, and by requiring scale symmetry invariance at the classical level\footnote{Quantum effects would contribute to some of the DEFT operators, modifying the overall Lagrangian coefficients. The explicit computation of the quantum contributions will not be discussed here as they are beyond the scope of constructing the DEFT Lagrangian.} by implementing Eq.~(\ref{ricetta}). Following Sect.~\ref{Sect:HEFT} and the description in App.~\ref{APP:Counting}, the DEFT Lagrangian can be written as the sum of two terms,
\be
\LL_\text{DEFT}\equiv \LL_{\chi}^{(0)} + \Delta \LL_{\chi}\,,
\label{DEFTLAG}
\ee
where the first term is very similar to $\LL_h^{(0)}$ of Eq.~(\ref{Lag0}):
\begin{align}
\LL_\chi^{(0)}=& -\dfrac{1}{4} \BBd\BBu -\dfrac{1}{4} \WWd^a W^{a\,\mu\nu}-\dfrac{1}{4} \GGd^\alpha \mathcal{G}^{\alpha\,\mu\nu}+\nn\\
&+\dfrac{1}{2}\de_\mu \chi\, \de^\mu \chi-\dfrac{v^2}{4}\left(\dfrac{\chi}{f}\right)^2\Tr(\V_\mu \V^\mu)-V(\chi)
+\nn\\
&+i\bar{Q}_L\slashed{D}Q_L+i\bar{Q}_R\slashed{D}Q_R+i\bar{L}_L\slashed{D}L_L
+i\bar{L}_R\slashed{D}L_R+\nn\\
&-\dfrac{v}{\sqrt2}\dfrac{\chi}{f}\left(\bar{Q}_L\U \cY_Q Q_R+\hc\right)+\label{LagD0}\\
&-\dfrac{v}{\sqrt2}\dfrac{\chi}{f}\left(\bar{L}_L\U \cY_L L_R+\hc\right)+\nn\\
&-\dfrac{g_s^2}{(4\pi)^2}\lambda_s\,\GGd^\alpha\, 
\tilde{\mathcal{G}}^{\alpha\,\mu\nu}\,.\nn
\end{align}
where
\begin{equation}
\cY_{Q}\equiv \diag \left(Y_{U}^{(0)}, Y_{D}^{(0)}\right)\,,\quad
\cY_{L}\equiv \diag\left(0, Y_{\ell}^{(0)}\right)\,.
\end{equation}
This Lagrangian contains the kinetic and mass terms for all the fields of the spectrum and leading interactions among SM gauge bosons and fermions and the dilaton. As for the HEFT, no interaction is  present at this order between the transverse components of the EW gauge bosons and the dilaton: here, however, in contrast to the HEFT, it does not follow any assumption, but it is simply due to the fact that the kinetic terms are already scale symmetry invariant and in consequence do not allow any additional insertion of $(\chi/f)$. The same holds for the kinetic terms for fermions and for the dilaton. On the contrary, EW gauge boson mass and Yukawa terms require the presence of $\chi$ in order to guarantee the scale symmetry invariance. 

Notice that, as for HEFT, a fine-tuning is present in order to correctly describe the scale of the SM gauge boson and fermion masses. This should not, however, be taken as a reason to not consider DEFT as a valid approach to describe EW and Higgs interactions: this is just an indication that the theory beyond this effective field theory should account for a mechanism which predicts the correct scales.

Finally, $V(\chi)$ contains non-derivative self-couplings of the conformal compensator $\chi$: $V(\chi)$ may contain scale symmetry breaking factors, including the dilaton mass, whose magnitude depends on the underlying theory considered. To avoid entering into details of specific realisation, the explicit form of $V(\chi)$ will not be discussed here and it will only be assumed that $V(\chi)$ is minimised by $\langle\chi\rangle=f$.

Rewriting $\LL_\chi^{(0)}$ in terms of the dilaton field $\sigma$ through Eq.~(\ref{sigmadef}), it follows that the dilaton kinetic term is not canonically normalised:
\be
\begin{aligned}
\frac{1}{2}\derp_\mu \chi\, \derp^\mu \chi 
&= \frac{1}{2} \derp_\mu \sigma \derp^\mu \sigma \, e^{2\sigma/f}\\
&= \frac{1}{2} \derp_\mu \sigma \derp^\mu \sigma \, \left(1+\dfrac{2\sigma}{f}+\dfrac{2\sigma^2}{f^2}+\ldots\right)\,,
\end{aligned}
\ee
where in the last equality the Taylor series expansion of the exponential has been explicitly written and dots stand for higher powers in $\sigma/f$. Performing the following redefinition on the dilaton field,
\be
\sigma=f\ln\left(\dfrac{\bar\chi}{f}+1\right)\,,
\label{Sigmaredef}
\ee
canonical normalised kinetic terms are recovered, where $\bc$ are the fluctuations of $\chi$ around its VEV. The resulting leading Lagrangian after this redefinition looks exactly as in Eq.~(\ref{LagD0}), except for a few terms, where $\chi$ insertions are substituted by insertions of $\bc+f$:
\begin{subequations}
\label{LD0afterRedef}
\begin{align}
\hspace{-1cm}\de_\mu \chi\, \de^\mu \chi\to& \de_\mu \bc\, \de^\mu \bc\,,\\
\hspace{-1cm}\left(\dfrac{\chi}{f}\right)^2\Tr(\V_\mu \V^\mu)\to&\left(1+\dfrac{\bc}{f}\right)^2\Tr(\V_\mu \V^\mu)\,,
\label{LD0afterRedefVmu}\\
\hspace{-1cm}\dfrac{\chi}{f}\bar{Q}_L\U \cY_Q Q_R\to&\left(1+\dfrac{\bc}{f}\right)\bar{Q}_L\U \cY_Q Q_R\,,\\
\hspace{-1cm}\dfrac{\chi}{f}\bar{L}_L\U \cY_L L_R\to&\left(1+\dfrac{\bc}{f}\right)\bar{L}_L\U \cY_L L_R\,.
\end{align}
\end{subequations}
It is interesting to note that Eq.~(\ref{LD0afterRedefVmu}), once assuming $f=v$, coincides with the SM result, that is Eq.~(\ref{FC}) for $a_C=1=b_C$ and the rest of coefficients vanishing. This well-known result (see for example Ref.~\cite{Goldberger:2008zz}) points out that only a partial unitarisation of the SM amplitudes occurs in the DEFT as long as $f>v$. The full unitarisation should then be accomplished by new degrees of freedom, that are expected to arise at the scale $\Lambda$, for example assuming an underlying strong dynamics.\\

The second term of the DEFT Lagrangian in Eq.~(\ref{DEFTLAG}) contains couplings of the dilaton that go beyond the SM-like ones of Eqs.~(\ref{LagD0}) and (\ref{LD0afterRedef}), and therefore could provide discriminating signals for shedding light on the Higgs nature. $\Delta\LL_\chi$ contains all the contributions that are generated from the 1-loop renormalisation. Restricting only on the CP-even bosonic couplings, $\Delta\LL_\chi$ can be written as the sum of distinct operators:
\begin{align}
\Delta\LL_\chi= & d_T \PP_T\FF_T(\chi)+d_B\PP_B \FF_B(\chi)+\nn\\
&+ d_W\PP_W\FF_W(\chi)+d_G\PP_G\FF_G(\chi) +\nn\\
&+d_{\Box H} \PP_{\Box H}\FF_{\Box H}(\chi)+d_{\Delta H} \PP_{\Delta H}\FF_{\Delta H}(\chi)\nn\\
&+d_{DH} \PP_{DH}\FF_{DH}(\chi)+\nn\\
&+d_{WWW}\PP_{WWW}\FF_{WWW}(\chi)+\label{DeltaLDilaton}\\
&+d_{GGG}\PP_{GGG}\FF_{GGG}(\chi)+d_{DB} \PP_{DB}\FF_{DB}(h)+\nn\\
&+d_{DW} \PP_{DW}\FF_{DW}(h)+d_{DG} \PP_{DG}\FF_{DG}(h)+\nn\\
&+\sum_{i=1}^{26} d_i\PP_i\FF_i(\chi)\,,\nn
\end{align}
where the parameters $d_i$ are free coefficients smaller than 1 -- in particular products of two or more operator coefficients $d_i$ will be neglected in the phenomenological analysis of the next sections. Moreover, the operators $\PP_i$ are similar to the list of operators $\cP_i$ in Eqs.(\ref{DeltaL})--(\ref{HEFTOperators}) for the HEFT, but with the proper modifications for the dilaton case discussed in this section. On the other side, the functions $\FF_i(\chi)$, differently from the functions $\cF(h)$ of the HEFT, are {\it not} arbitrary functions of the dilaton field, but depend on the specific operator considered.

In the first line of Eq.~(\ref{DeltaLDilaton}),
\be
\begin{split}
\PP_T&\FF_T(\chi)=\frac{v^2}{4} \tr(\T\V_\mu)\tr(\T\V^\mu)\dfrac{\chi^2}{f^2}\\
&\to\frac{v^2}{4} \tr(\T\V_\mu)\tr(\T\V^\mu)\left(1+\dfrac{\bc}{f}\right)^2\,,
\end{split}
\ee
where the operator in the first (second) line is written before (after) moving to the basis of canonical kinetic term for the dilaton, Eq.~(\ref{Sigmaredef}). Moreover, the terms in the brackets correspond to the function $\FF_T(\chi)$. The $\bc$-independent term provides a tree-level contribution to the EW precision parameter  $T$, exactly as for the operator $\cP_T$ in Eq.~(\ref{P_T}), and therefore the coefficient $d_T$ is constrained to be at most at the percent level, such as $c_T$. This could indicate that the conformal sector is custodial preserving: if this is the case, all the custodial breaking operators are expected to be suppressed. 

The four operators $\PP_{B}$, $\PP_{W}$, $\PP_{G}$ and $\PP_{1}$ describe the interaction of the dilaton with the transverse components of the gauge bosons:
\be
\begin{aligned}
\PP_{B} \FF_B(\chi)&=-\frac{1}{4}\BBd \BBu \dfrac{\bc}{f} \\
\PP_{W} \FF_W(\chi)&=-\frac{1}{4}\WWd^a W^{a\,\mu\nu} \dfrac{\bc}{f} \\
\PP_G \FF_G(\chi)&= -\frac{1}{4}\GGd^\alpha\, \mathcal{G}^{\alpha\,\mu\nu} \dfrac{\bc}{f}\\
\PP_{1}\FF_1(\chi)&=B_{\mu\nu} \Tr\left(\T\,\W^{\mu\nu}\right)\dfrac{\bc}{f}\,,
\end{aligned}
\label{DilatonGaugeBosons}
\ee
where the last factor $\bc/f$ in each operator represents the functions $\FF_i(\chi)$.
From the point of view of the scale invariance, insertion of $\bc/f$ would not be necessary and therefore could be ignored in the most conservative approach. However, 1-loop contributions to these couplings may arise: for example, if the conformal sector is charged under the SM gauge group, the scale symmetry is anomalous and the SM running does induce these couplings at 1-loop level. Although no specific assumption on the underlying dynamics is taken here, the choice in Eq.~(\ref{DilatonGaugeBosons}) is of considering generic couplings that may encode these loop contributions. Explicit computations of the dilaton interactions to massless gauge bosons have been computed for example in Ref.~\cite{Goldberger:2008zz}. The same reasoning can be easily extended to couplings to massive gauge bosons, $WW\bc$ and $ZZ\bc$, and to the $Z\gamma\bc$.

The three operators $\PP_{\Box \chi}$, $\PP_{\Delta \chi}$ and $\PP_{D\chi}$ have a similar structure to the one of the HEFT operators $\cP_{\Box H}$, $\cP_{\Delta H}$ and $\cP_{DH}$ defined in Eq.~(\ref{SelfHiggsOperators}): 
\begin{align}
&&\begin{split}
\PP_{\Box \chi}&\FF_{\Box \chi}(\chi)= \dfrac{1}{(4\pi)^2\chi^2}\left(\Box  \chi \right)^2\\
\to& \frac{1}{\Lambda^2}\left(\Box  \bc \right)^2\left(1-\frac{2 \bc}{f}+\frac{3 \bc^2}{f^2}+\ldots\right)
\end{split}\nn\\
&&
\PP_{\Delta \chi}&\FF_{\Delta \chi}(\chi)=\dfrac{1}{(4\pi)^2\chi^3}\left(\derp_\mu \chi\,\derp^\mu \chi\right) \Box  \chi
\label{SelfDilatonOperators}\\
&&\to&\frac{4\pi}{\Lambda^3}\left(\derp_\mu \bc\derp^\mu \bc\right)\left(\Box  \bc \right)\left(1-\frac{3 \bc}{f}+\frac{6 \bc^2}{f^2}+\ldots\right)\nn\\
&&\begin{split}
\PP_{D\chi}&\FF_{D\chi}(\chi) =\dfrac{1}{(4\pi)^2\chi^4} \left(\derp_\mu \chi\,\derp^\mu \chi \right)^2 \\
\to& \dfrac{(4\pi)^2}{\Lambda^4}\left(\derp_\mu \bc\,\derp^\mu \bc \right)^2\left(1-\frac{4 \bc}{f}+\frac{10 \bc^2}{f^2}+\ldots\right)\,,
\end{split}\nn
\end{align}
where $\Lambda=4\pi f$ has been used and the last factors inside the brackets represent the functions $\FF_i(\chi)$. 

A first difference between Eqs.~(\ref{SelfHiggsOperators}) and (\ref{SelfDilatonOperators}) is the absence of an equivalent operator to $\cP_H$: being $\de_\mu \chi \de^\mu \chi$ already of $\dim=4$, no additional $\chi^n$ insertion is necessary for scale invariance; moreover, differently from what discussed for the operators in Eq.~(\ref{DilatonGaugeBosons}), no physical\footnote{Any non-derivative polynomial of $\chi$ could be redefined away by a redefinition of the field $\chi$. This would translate in a modification, proportional to the corresponding operator coefficient, of all the other $\chi$ couplings in the Lagrangian. As the product of two or more operator coefficients are neglected, this redefinition would not have any impact. See App.~B in Ref.~\cite{Brivio:2016fzo} for further details.} 1-loop correction affects this operator. This, however, does not lead to any physical difference between the HEFT and the DEFT, as the combination $\cP_H\cF_H(h)$ can be safely reabsorbed through an $h$ redefinition~\cite{Giudice:2007fh}.

A second important difference appears in the functions $\cF_i(h)$ and $\FF_i(\chi)$: the first ones are completely generic polynomials of $h/v$ in the HEFT; for the dilaton case, only specific numerical coefficients multiply the distinct factors $\bc/f$. When the experimental sensitivity on the involved observables will be high enough, the identification of these numerical coefficients could tell a Higgs-like dilaton from other possibilities, as indeed they are different in the SMEFT case, where $\cF_i(h)$ can only be powers of $(1+h/v)^2$, or in CH models where $\cF_i(h)$ are trigonometric functions of $h/f$: this will be further discussed in the next sections.

The normalisation of the operators in Eq.~(\ref{SelfDilatonOperators}) has been fixed so as to match the one of the operators in Eq.~(\ref{SelfHiggsOperators}): although this is not necessary from the EFT point of view, with this choice one recovers the Naive Dimensional Analysis normalisation, which tells that the theory enters into a strong interacting regime when $d_i=1$. Indeed, a theory enters into a strong interacting regime when loop corrections to the Wilson coefficient of a generic operator turn out to be (at least) as important as the initial value of the Wilson coefficient itself. The Naive Dimensional Analysis normalisation used in Eq.~(\ref{DeltaL}) for HEFT identifies this phase transition with the condition of having operator coefficient equal or larger than $1$. A shortcut to ensure this condition for the dilaton case is to profit of the similarities between the DEFT and the HEFT Lagrangians, when identifying the scale $\Lambda$ as the same scale in both frameworks. The numerical factors in the operators in Eq.~(\ref{SelfDilatonOperators}) -- and in the equations which follow -- has been fixed in this way.

The five operators $\PP_{WWW}$, $\PP_{GGG}$, $\PP_{DB}$, $\PP_{DW}$ and $\PP_{DG}$ in the last lines of Eq.~(\ref{DeltaLDilaton}) are defined as
\begin{widetext}
\be
\begin{aligned}
\PP_{WWW}&\FF_{WWW}(\chi) =\dfrac{\varepsilon_{abc}}{4\pi\chi^2}W_\mu^{a\nu}W_\nu^{b\rho} W_{\rho}^{c\mu}\\
\to&\dfrac{4\pi\varepsilon_{abc}}{\Lambda^2}W_\mu^{a\nu}W_\nu^{b\rho} W_{\rho}^{c\mu}\left(1-\frac{2 \bc}{f}+\frac{3 \bc^2}{f^2}+\ldots\right)\\
\PP_{GGG}&\FF_{GGG}(\chi)=\dfrac{f_{\alpha\beta\gamma}}{4\pi\chi^2}{\mathcal G}_\mu^{\alpha\nu}{\mathcal G}_\nu^{\beta\rho} {\mathcal G}_{\rho}^{\gamma\mu}\\
\to&\dfrac{4\pi f_{\alpha\beta\gamma}}{\Lambda^2}{\mathcal G}_\mu^{\alpha\nu}{\mathcal G}_\nu^{\beta\rho} {\mathcal G}_{\rho}^{\gamma\mu}\left(1-\frac{2 \bc}{f}+\frac{3 \bc^2}{f^2}+\ldots\right)
\end{aligned}\quad
\begin{aligned}
\PP_{DB}&\FF_{DB}(\chi)=\dfrac{1}{(4\pi)^2\chi^2}\left(\derp^\mu B_{\mu\nu}\right)\left(\derp_\rho B^{\rho\nu}\right)\\
\to&\dfrac{1}{\Lambda^2}\left(\derp^\mu B_{\mu\nu}\right)^\alpha\left(\derp_\rho B^{\rho\nu}\right)^\alpha\left(1-\frac{2 \bc}{f}+\frac{3 \bc^2}{f^2}+\ldots\right)\\
\PP_{DW}&\FF_{DW}(\chi)=\dfrac{1}{(4\pi)^2\chi^2}\left(\cD^\mu W_{\mu\nu}\right)^a\left(\cD_\rho W^{\rho\nu}\right)^a\\
\to&\dfrac{1}{\Lambda^2}\left(\cD^\mu W_{\mu\nu}\right)^a\left(\cD_\rho W^{\rho\nu}\right)^a\left(1-\frac{2 \bc}{f}+\frac{3 \bc^2}{f^2}+\ldots\right)\\
\PP_{DG}&\FF_{DG}(\chi)=\dfrac{1}{(4\pi)^2\chi^2}\left(\cD^\mu G_{\mu\nu}\right)^\alpha\left(\cD_\rho G^{\rho\nu}\right)^\alpha\\
\to&\dfrac{1}{\Lambda^2}\left(\cD^\mu G_{\mu\nu}\right)^\alpha\left(\cD_\rho G^{\rho\nu}\right)^\alpha\left(1-\frac{2 \bc}{f}+\frac{3 \bc^2}{f^2}+\ldots\right)\,,
\end{aligned}
\ee
\end{widetext}
where the last factors inside the brackets represent the functions $\FF_i(\chi)$.

The remaining terms entering the Lagrangian $\Delta\LL_\chi$ are four-derivative operators and are defined in the following Eq.~(\ref{DEFTOperators}). 

Part of the operators required the insertion of the conformal regulator in order to recover scale invariance: these terms have been written both in terms of $\chi$, to make explicit the scale invariance although the dilaton kinetic terms are not canonical yet, and in terms of $\bc$, that is in the canonical kinetic term basis. The rest of operators are naturally of canonical dimension $\dim=4$ and thus scale invariant by themselves. An explicit $\chi$ dependence has not been added to these operators, differently from what done for those in Eq.~(\ref{DilatonGaugeBosons}). The reason resides in the fact that the operators in Eq.~(\ref{DEFTOperators}) are generated already at the first subleading order: as any $\chi$ insertion would be the consequence of an additional loop effect, this would be equivalent to a global two-loop contribution that should not be considered at this level of the expansion.

As for the HEFT operators listed in Eq.~(\ref{HEFTOperators}), those containing the scalar chiral field $\T$, not in association to the gauge field strength $B_{\mu\nu}$, represent sources of custodial symmetry breaking beyond those of the SM: specifically, they are $\PP_{12}-\PP_{19}$ and $\PP_{21}-\PP_{26}$. If the conformal dynamics behind the DEFT Lagrangian is custodial preserving, then all these operators, together with $\PP_T$, are expected to be further suppressed.

Finally, the list of operators entering $\Delta\LL_\chi$ reads as follows:
\begin{widetext}
\be
\begin{aligned}
\begin{aligned}
\PP_{2}\FF_2(\chi)  &=\dfrac{i}{4\pi}B_{\mu\nu} \Tr\left(\T\left[\V^\mu,\V^\nu\right]\right) \\
\PP_{3}\FF_3(\chi)  &= \dfrac{i}{4\pi}\Tr\left(\W_{\mu\nu}\left[\V^\mu,\V^\nu\right]\right)\\
\PP_{4}\FF_4(\chi)  &= \dfrac{i}{4\pi\chi}B_{\mu\nu}\Tr(\T\V^\mu)\,\derp^\nu \chi\\
\to\dfrac{i}{\Lambda}&B_{\mu\nu}\Tr(\T\V^\mu)\,\derp^\nu \bc\left(1-\frac{\bc}{f}+\frac{\bc^2}{f^2}+\ldots\right)\\
\PP_{5}\FF_5(\chi)  &= \dfrac{i}{4\pi\chi}\Tr(\W_{\mu\nu}\V^\mu)\,\derp^\nu \chi \\
\to \dfrac{i}{\Lambda}&\Tr(\W_{\mu\nu}\V^\mu)\,\derp^\nu \bc\left(1-\frac{\bc}{f}+\frac{\bc^2}{f^2}+\ldots\right) \\
\PP_{6}\FF_6(\chi) &=\dfrac{1}{(4\pi)^2}\left(\Tr\left(\V_\mu\,\V^\mu\right)\right)^2 \\
\PP_{7}\FF_7(\chi) &=\dfrac{1}{(4\pi)^2\chi}\Tr\left(\V_\mu\,\V^\mu\right)\derp_\nu\derp^\nu \chi\\
\to\dfrac{1}{4\pi\Lambda}&\Tr\left(\V_\mu\,\V^\mu\right)\derp_\nu\derp^\nu \bc\left(1-\frac{\bc}{f}+\frac{\bc^2}{f^2}+\ldots\right)\\
\PP_{8}\FF_8(\chi) &=\dfrac{1}{(4\pi)^2\chi^2}\Tr\left(\V_\mu\,\V_\nu\right)\derp^\mu \chi\,\derp^\nu \chi\\
\to \dfrac{1}{\Lambda^2}&\Tr\left(\V_\mu\,\V_\nu\right)\derp^\mu \bc\,\derp^\nu \bc\left(1-\frac{2 \bc}{f}+\frac{3 \bc^2}{f^2}+\ldots\right)\\
\PP_{9}\FF_9(\chi) &=\dfrac{1}{(4\pi)^2}\Tr\left((\cD_\mu\V^\mu)^2 \right)\\
\PP_{10}\FF_{10}(\chi) &=\dfrac{1}{(4\pi)^2\chi}\Tr(\V_\nu \,\cD_\mu\V^\mu)\,\derp^\nu \chi\\
\to \dfrac{1}{4\pi\Lambda}&\Tr(\V_\nu \,\cD_\mu\V^\mu)\,\derp^\nu\bc\left(1-\frac{\bc}{f}+\frac{\bc^2}{f^2}+\ldots\right)\\
\PP_{11}\FF_{11}(\chi) &=\dfrac{1}{(4\pi)^2}\left(\Tr\left(\V_\mu\,\V_\nu\right)\right)^2\\
\PP_{12}\FF_{12}(\chi) &=\left(\Tr\left(\T \W_{\mu\nu}\right)\right)^2\\
\PP_{13}\FF_{13}(\chi) &=\dfrac{i}{4\pi}\Tr\left(\T \W_{\mu\nu}\right)\Tr\left(\T\left[\V^\mu,\V^\nu\right]\right)\\
\PP_{14}\FF_{14}(\chi) &=\dfrac{\varepsilon_{\mu\nu\rho\lambda}}{4\pi}\Tr\left(\T \V^\mu\right)\Tr\left(\V^\nu \W^{\rho\lambda}\right)\\
\PP_{15}\FF_{15}(\chi) &= \dfrac{1}{(4\pi)^2}\Tr(\T\,\cD_\mu\V^\mu)\,\Tr(\T\,\cD_\nu\V^\nu)
\end{aligned}
\qquad
\begin{aligned}
\PP_{16}\FF_{16}(\chi) & = \dfrac{1}{(4\pi)^2}\Tr([\T \,,\V_\nu]\,\cD_\mu \V^\mu) \, \Tr(\T\V^\nu)\\
\PP_{17}\FF_{17}(\chi) &= \dfrac{i}{4\pi\chi}\Tr(\T \W_{\mu\nu})\Tr(\T\V^\mu)\,\derp^\nu \chi\\
\to\dfrac{i}{\Lambda}&\Tr(\T \W_{\mu\nu})\Tr(\T\V^\mu)\,\derp^\nu \bc\left(1-\frac{\bc}{f}+\frac{\bc^2}{f^2}+\ldots\right)\\
\PP_{18}\FF_{18}(\chi) &=\dfrac{1}{(4\pi)^2\chi}\Tr(\T\,[\V_\mu,\V_\nu])\Tr(\T\V^\mu)\,\derp^\nu\chi\\
\to \dfrac{1}{4\pi\Lambda}&\Tr(\T\,[\V_\mu,\V_\nu])\Tr(\T\V^\mu)\,\derp^\nu \bc\left(1-\frac{\bc}{f}+\frac{\bc^2}{f^2}+\ldots\right)\\
\PP_{19}\FF_{19}(\chi) &=\dfrac{1}{(4\pi)^2\chi}\Tr(\T\,\cD_\mu\V^\mu)\Tr(\T\V_\nu)\,\derp^\nu \chi\\
\to\dfrac{1}{4\pi\Lambda}&\Tr(\T\,\cD_\mu\V^\mu)\Tr(\T\V_\nu)\,\derp^\nu \bc\left(1-\frac{\bc}{f}+\frac{\bc^2}{f^2}+\ldots\right)\\
\PP_{20}\FF_{20}(\chi) &=\dfrac{1}{(4\pi)^2\chi^2}\Tr\left(\V_\mu\,\V^\mu\right)\derp_\nu \chi\, \derp^\nu \chi\\
\to\dfrac{1}{\Lambda^2}&\Tr\left(\V_\mu\,\V^\mu\right)\derp_\nu \bc\, \derp^\nu \bc\left(1-\frac{2 \bc}{f}+\frac{3 \bc^2}{f^2}+\ldots\right)\\
\PP_{21}\FF_{21}(\chi) &=\dfrac{1}{(4\pi)^2\chi^2}\left(\Tr\left(\T\,\V_\mu\right) \right)^2 \derp_\nu \chi\, \derp^\nu \chi\\
\to\dfrac{1}{\Lambda^2}&\left(\Tr\left(\T\,\V_\mu\right) \right)^2 \derp_\nu \bc\, \derp^\nu \bc\left(1-\frac{2 \bc}{f}+\frac{3 \bc^2}{f^2}+\ldots\right)\\
\PP_{22}\FF_{22}(\chi) &=\dfrac{1}{(4\pi)^2\chi^2}\Tr\left(\T\V_\mu\right)\Tr\left(\T\V_\nu\right)\,\derp^\mu \chi\,\derp^\nu \chi\\
\to\dfrac{1}{\Lambda^2}&\Tr\left(\T\V_\mu\right)\Tr\left(\T\V_\nu\right)\,\derp^\mu \bc\,\derp^\nu \bc\left(1-\frac{2 \bc}{f}+\frac{3 \bc^2}{f^2}+\ldots\right)\\
\PP_{23}\FF_{23}(\chi) &=\dfrac{1}{(4\pi)^2} \Tr\left(\V_\mu\V^\mu\right)\left(\Tr\left(\T\V_\nu\right)\right)^2\\
\PP_{24}\FF_{24}(\chi) &=\dfrac{1}{(4\pi)^2}\Tr\left(\V_\mu\V_\nu\right)\Tr\left(\T\V^\mu\right)\Tr\left(\T\V^\nu\right) \\
\PP_{25}\FF_{25}(\chi) &=\dfrac{1}{(4\pi)^2\chi}\left(\Tr\left(\T\,\V_\mu\right) \right)^2 \derp_\nu\derp^\nu \chi\\
\to&\dfrac{1}{4\pi\Lambda}\left(\Tr\left(\T\,\V_\mu\right) \right)^2 \derp_\nu\derp^\nu \bc\left(1-\frac{\bc}{f}+\frac{\bc^2}{f^2}+\ldots\right)\\
\PP_{26}\FF_{26}(\chi) &=\dfrac{1}{(4\pi)^2}\left(\Tr\left(\T\V_\mu\right)\Tr\left(\T\V_\nu\right)\right)^2\,,
\end{aligned}
\end{aligned}
\label{DEFTOperators}
\ee
\end{widetext}
where the functions $\FF_i(\chi)$ correspond to the the last factors inside the brackets, when present, otherwise are equal to 1.\footnote{Adding an explicit $\FF_i(\chi)$ function to those operators that do not have it would be equivalent to a two-loop effect, which should not be considered at the expansion order of the DEFT Lagrangian presented here.}

Eq.~(\ref{DEFTOperators}) completes the DEFT operator basis describing CP-even bosonic couplings of the SM EW sector and the Higgs-like dilaton. Any other non-fermionic CP-even operator that can be constructed can be written in terms of the operators of this basis, by using partial integration, Bianchi identity and $SU(2)_L$-algebra.

%
%
\section{Disentangling a Higgs-Like Dilaton}
\label{Sect:Comparison}

A Higgs-like dilaton is typically considered to have the same couplings as the SM Higgs, except for the non-derivative self-couplings entering the scalar potential. The previous section illustrates that this is the case for a restricted group of operators, while there are several other couplings which differ from the SM context. This section is dedicated to discussing these differences considering the SMEFT Lagrangian, whose specific limit is the SM, and the minimal $SO(5)/SO(4)$ CH model~\cite{Agashe:2004rs}, making use of the HEFT Lagrangian defined in Sect.~\ref{Sect:HEFT} as a common background. The explicit operators basis for the SMEFT Lagrangian can be found in App.~\ref{APP:SMEFT}, while the one for the minimal $SO(5)/SO(4)$ CH model is in App.~\ref{APP:CH}.\\

The first coupling that can be considered for this comparison is the one responsible for the $W^{\pm}$ and $Z$ masses. In the SMEFT up to $\dim=6$ expansion order, there are four relevant operators:
\be
\DL_\mu \Phi^\dag\DL^\mu\Phi+\dfrac{c_{\Phi,1}}{\Lambda^2}\cO_{\Phi,1}+\dfrac{c_{\Phi,2}}{\Lambda^2}\cO_{\Phi,2}+\dfrac{c_{\Phi,4}}{\Lambda^2}\cO_{\Phi,4}\,,
\ee
where the explicit definition of the operators $\cO_{\Phi,i}$ can be found in Eq.~(\ref{LinearOpsGauge}). As described in App.~\ref{APP:SMEFT}, $\cO_{\Phi,1}$ and $\cO_{\Phi,2}$ contribute to the EW gauge boson masses only indirectly, through the transformation to move to the basis of canonical Higgs kinetic terms. In the latter basis, projecting into the HEFT basis, $\cF_C(h)$ is given by
\begin{align}
\cF^\text{SMEFT}_C(h)=&\left[\left(1+\frac{(4\pi)^2}{2}\frac{v^2}{\Lambda^2}c_{\Phi4}\right)+\right.\\
&+2\frac{h}{v}\left(1-\frac{(4\pi)^2}{4}\frac{v^2}{\Lambda^2}\left(c_{\Phi1}+2c_{\Phi2}-3c_{\Phi4}\right)\right)+\nn\\
&+\frac{h^2}{v^2}\left(1-(4\pi)^2\frac{v^2}{\Lambda^2}\left(c_{\Phi1}+2c_{\Phi2}-2c_{\Phi4}\right)\right)+\nn\\
&\left.+\ldots\right]\,,\nn
\end{align}
where the dots stand for terms proportional to $h^3$ and $h^4$. When the coefficients $c_{\Phi,i}=0$, then the SM case is recovered.

For the minimal $SO(5)/SO(4)$ CH model and considering up to the NLO expansion order, only one contribution is present:
\be
-\frac{f^2}{4}\Tr\left(\Vt_\mu\Vt^\mu\right)=-\dfrac{v^2}{4}\Tr(\V_\mu \V^\mu)\left(\dfrac{4f^2}{v^2}\sin^2\frac{\varphi}{2f}\right)\,,
\ee
where $\Vt$ is the generalised vector chiral field in the $SO(5)$ representation, sibling of the vector chiral field $\V$ of the EW chiral Lagrangian, and $\varphi=h+\vh$, with $\sin^2(\vh/2f)=v^2/4f^2$. The $\cF_C(h)$ function for this case can easily be read from the previous expression:
\be
\cF^\text{CH}_C(h)=\dfrac{4f^2}{v^2}\sin^2\frac{h+\vh}{2f}\,.
\ee

These results should be compared with the equivalent contributions in the dilaton context, Eqs.~(\ref{LagD0}) and (\ref{LD0afterRedefVmu}): using a similar notation as for the HEFT, one can write
\be
\FF^\text{DEFT}_C(\chi)=\left(1+\dfrac{\bc}{f}\right)^2\,.
\label{FCDilaton}
\ee

All the three descriptions expect deviations from the SM predictions for the Higgs couplings to longitudinal components of the massive gauge bosons, and they differ from each other. The DEFT is the only one presenting up to quadratic powers of $h$ and with the same numerical factors as in the SM, even if it has a $v/f$ suppressing factor. EW precision tests as well as Tevatron and LHC data on production cross sections favours values for $v/f\lesssim 1$, indicating that the scale of the scale symmetry breaking is close to the EWSB one. A recent analysis on this aspect can be found in Ref.~\cite{Bellazzini:2012vz}. Notice that this bound on the scale $f$ is characteristic of the dilaton context; in the CH framework, $f$ is instead distinct from $v$ and present collider bounds suggest that $f\gtrsim4 v$.\\

A second group of relevant couplings that deserves a dedicated discussion is represented by the Yukawa interactions. In general, deviations from the SM predictions are expected in all the frameworks. Here, however, the focus is on the bosonic Lagrangians only and no genuine fermionic operator is considered. In consequence, the only modification to the SM Yukawa interactions that can arise is due to the transformation to move to the basis of canonical kinetic terms, which in particular occurs in both the SMEFT and the DEFT Lagrangians. Then, the functions multiplying the fermionic bilinear $(v/\sqrt2)\bar{\psi}_L\U Y^{(0)}_\psi\psi_R$ read:
\begin{align}
\cF^\text{SMEFT}_{U,D,\ell}(h)=&1+\frac{h}{v}\left(1-\frac{(4\pi)^2}{12}\left(c_{\Phi1}+2c_{\Phi2}+c_{\Phi4}\right)\times\right.\nn\\
&\hspace{3cm}\left.\times\dfrac{3v^2+3vh+h^2}{\Lambda^2}\right)\nn\\
\cF^\text{CH}_{U,D,\ell}(h)=&\sin\frac{h+\vh}{f}\label{FYDilaton}\\
\FF^\text{DEFT}_{U,D,\ell}(\chi)=&1+\dfrac{\bc}{f}\,.\nn
\end{align}
For the minimal CH model, the $\cF^\text{CH}_{U,D,\ell}(h)$ expression follows the result of the original formulation in Ref.~\cite{Agashe:2004rs}. As for the couplings with the longitudinal components of the EW gauge bosons, only the DEFT setup presents the same powers of $h$ and numerical factors as in the SM, except for the $v/f$ factor. Once considering that $v\lesssim f$ as indicated from the dilaton couplings to the longitudinal components of the EW gauge bosons, the dilaton couplings to fermions tend to align to the SM predictions. It is worth mentioning that these couplings are typically modified by the anomalous dimensions of the SM fermions and of the fields responsible of the EWSB, thus allowing for small deviations from the SM predictions (eventually, flavour changing neutral current contributions may arise). This depends on the specific fermion context considered in the underlying theory, examples of which can be found in Refs.~\cite{Vecchi:2010gj,Bellazzini:2012vz}.\\

\begin{table*}
{\small
\renewcommand{\arraystretch}{2}
\begin{tabular}{|>{$}c<{$}|>{$}c<{$}|>{$}c<{$}||>{$}c<{$}|>{$}c<{$}|}
\hline
c_i\cF_i(h)& \text{SMEFT}\,\,\, \dim\leq6  & SO(5)/SO(4) & d_i\FF_i(\chi) &\text{DEFT}  \\ 
\hline
\hline
c_B\cF_B(h)
& 2(4\pi)^2\frac{v^2}{\Lambda^2}c_{BB}\left(1+\frac{h}{v}\right)^2
& -4\ct_{B\Sigma}\cos^2\frac{\varphi}{2f}
& d_{B}\FF_{B}(\chi)
& d_B\frac{\bc}{f}
\\
c_W\cF_W(h)
& \frac{(4\pi)^2}{2}\frac{v^2}{\Lambda^2}c_{WW}\left(1+\frac{h}{v}\right)^2
& -4\ct_{W\Sigma}\cos^2\frac{\varphi}{2f}
& d_{W}\FF_{W}(\chi)
& d_W\frac{\bc}{f}
\\
c_G\cF_G(h)
& 2(4\pi)^2\frac{v^2}{\Lambda^2}c_{GG}\left(1+\frac{h}{v}\right)^2
& *
& d_{G}\FF_{G}(\chi)
& d_G\frac{\bc}{f}
\\
c_{\Box H}\cF_{\Box H}(h)
& \frac{1}{2}c_{\Box \Phi}
& -2\ct_6
& d_{\Box\chi}\FF_{\Box\chi}(\chi)
& d_{\Box\chi}\left(1-\frac{2 \bc}{f}+\frac{3 \bc^2}{f^2}+\ldots\right)
\\
c_{\Delta H}\cF_{\Delta H}(h)
& -
& -
& d_{\Delta \chi}\FF_{\Delta\chi}(\chi)
& d_{\Delta \chi}\left(1-\frac{3 \bc}{f}+\frac{6 \bc^2}{f^2}+\ldots\right)
\\
c_{D H}\cF_{D H}(h)
& -
& 4\left(\ct_4+\ct_5\right)
& d_{D\chi}\FF_{D\chi}(\chi)
& d_{D\chi}\left(1-\frac{4 \bc}{f}+\frac{10 \bc^2}{f^2}+\ldots\right)
\\
c_{WWW}\cF_{WWW}(h)
& c_{3W}
& -
& d_{WWW}\FF_{WWW}(\chi)
& d_{WWW}\left(1-\frac{2 \bc}{f}+\frac{3 \bc^2}{f^2}+\ldots\right)
\\
c_{GGG}\cF_{GGG}(h)
& c_{3G}
& -
& d_{GGG}\FF_{GGG}(\chi)
& d_{GGG}\left(1-\frac{2 \bc}{f}+\frac{3 \bc^2}{f^2}+\ldots\right)
\\
c_{DB}\cF_{DB}(h)
& c_{DB}
& -
& d_{DB}\FF_{DB}(\chi)
& d_{DB}\left(1-\frac{2 \bc}{f}+\frac{3 \bc^2}{f^2}+\ldots\right)
\\
c_{DW}\cF_{DW}(h)
& c_{DW}
& -
& d_{DW}\FF_{DW}(\chi)
& d_{DW}\left(1-\frac{2 \bc}{f}+\frac{3 \bc^2}{f^2}+\ldots\right)
\\
c_{DG}\cF_{DG}(h)
& c_{DG}
& -
& d_{DG}\FF_{DG}(\chi)
& d_{DG}\left(1-\frac{2 \bc}{f}+\frac{3 \bc^2}{f^2}+\ldots\right)
\\
c_1\cF_1(h)
& \frac{(4\pi)^2}{8}\frac{v^2}{\Lambda^2}c_{BW}\left(1+\frac{h}{v}\right)^2
& \ct_1\sin^2\frac{\varphi}{2f}
& d_{1}\FF_{1}(\chi)
& d_{1}\frac{\bc}{f}
\\
c_2\cF_2(h)
& \frac{(4\pi)^2}{8}\frac{v^2}{\Lambda^2}c_B\left(1+\frac{h}{v}\right)^2
& \ct_2\sin^2\frac{\varphi}{2f}
& d_{2}\FF_{2}(\chi)
& d_{2}
\\
c_3\cF_3(h)
& \frac{(4\pi)^2}{4}\frac{v^2}{\Lambda^2}c_W\left(1+\frac{h}{v}\right)^2
& 2\ct_3\sin^2\frac{\varphi}{2f}
& d_{3}\FF_{3}(\chi)
& d_{3}
\\
c_4\cF_4(h)
& \frac{4\pi}{2}\frac{v}{\Lambda}c_B\left(1+\frac{h}{v}\right)
& \ct_2\sin\frac{\varphi}{f}
& d_{4}\FF_{4}(\chi)
& d_{4}\left(1-\frac{\bc}{f}+\frac{\bc^2}{f^2}+\ldots\right)
\\
c_5\cF_5(h)
& -4\pi\frac{v}{\Lambda}c_W\left(1+\frac{h}{v}\right)
& -2\ct_3\sin\frac{\varphi}{f}
& d_{5}\FF_{5}(\chi)
& d_{5}\left(1-\frac{\bc}{f}+\frac{\bc^2}{f^2}+\ldots\right)
\\
c_6\cF_6(h)
& \frac{(4\pi)^2}{8}\frac{v^2}{\Lambda^2}c_{\Box \Phi}\left(1+\frac{h}{v}\right)^2
& 16\ct_4\sin^4\frac{\varphi}{2f}-\dfrac{1}{2}\ct_6\sin^2\frac{\varphi}{f} 
& d_{6}\FF_{6}(\chi)
& d_{6}
\\
c_7\cF_7(h)
& \frac{4\pi}{2}\frac{v}{\Lambda}c_{\Box \Phi}\left(1+\frac{h}{v}\right)
& -2\ct_6\sin\frac{\varphi}{f} 
& d_{7}\FF_{7}(\chi)
& d_{7}\left(1-\frac{\bc}{f}+\frac{\bc^2}{f^2}+\ldots\right)
\\
c_8\cF_8(h)
& -c_{\Box \Phi}
& -16\ct_5\sin^2\frac{\varphi}{2f}+4\ct_6\cos^2\frac{\varphi}{2f} 
& d_{8}\FF_{8}(\chi)
& d_{8}\left(1-\frac{2\bc}{f}+\frac{3\bc^2}{f^2}+\ldots\right)
\\
c_9\cF_9(h)
& -\frac{(4\pi)^2}{4}\frac{v^2}{\Lambda^2}c_{\Box \Phi}\left(1+\frac{h}{v}\right)^2
& 4\ct_6\sin^2\frac{\varphi}{2f} 
& d_{9}\FF_{9}(\chi)
& d_{9}
\\
c_{10}\cF_{10}(h )
& -4\pi\frac{v}{\Lambda}c_{\Box \Phi}\left(1+\frac{h}{v}\right)
& 4\ct_6\sin\frac{\varphi}{f} 
& d_{10}\FF_{10}(\chi)
& d_{10}\left(1-\frac{\bc}{f}+\frac{\bc^2}{f^2}+\ldots\right)
\\ 
c_{11}\cF_{11}(h)
& -
& 16\ct_5\sin^4\frac{\varphi}{2f} 
& d_{11}\FF_{11}(\chi)
& d_{11}
\\
c_{20}\cF_{20}(h)
& -
& -16\ct_4\sin^2\frac{\varphi}{2f} 
& d_{20}\FF_{20}(\chi)
& d_{20}\left(1-\frac{2\bc}{f}+\frac{3\bc^2}{f^2}+\ldots\right)
\\\hline
\end{tabular}}
\caption{
\it Explicit expressions for the products $c_{i}\cF_{i}(h)$ and $d_{i}\FF_{i}(\chi)$ of the low-energy custodial preserving operators. The ``$-$'' entries indicate the absence of contributions at the order considered to the corresponding low-energy operator. The ``*'' symbol indicates that couplings to gluons in CH models are expected due to fermionic loops (see App.~\ref{APP:CH}). 
}
\label{tableComparisonsCustCons}
\end{table*}

The rest of the couplings and the comparison among the different contributions from SMEFT, the minimal $SO(5)/SO(4)$ CH model and DEFT are summarised in Tab.~\ref{tableComparisonsCustCons}, restricting to the only custodial preserving contributions. It is worthwhile to make a comment about the couplings to the transverse components of the gauge bosons: as mentioned below Eq.~(\ref{DilatonGaugeBosons}), these interactions are not strictly required from scale symmetry invariance; however, they are considered here to encode possible 1-loop level contributions that may arise, for example if the conformal sector is charged under the SM gauge symmetry. Consequently, the coefficients $d_B$, $d_W$ and $d_G$ are expected to be smaller than 1. Considering that $v\lesssim f$ as indicated from collider data on the dilaton couplings to the longitudinal components of the SM gauge bosons, the suppression of these coefficients is a welcome byproduct of the DEFT construction, in order to reconcile this effective description with LHC data. 

\begin{table*}[ht!]
{\small
\renewcommand{\arraystretch}{2}
\begin{tabular}{|>{$}c<{$}|>{$}c<{$}||>{$}c<{$}|>{$}c<{$}||>{$}c<{$}|>{$}c<{$}|}
\hline
d_i\FF_i(\chi)& \text{DEFT}  & d_i\FF_i(\chi) &\text{DEFT} & d_i\FF_i(\chi) &\text{DEFT}  \\ 
\hline
\hline
d_{12}\FF_{12}(\chi)
& d_{12}
& d_{17}\FF_{17}(\chi)
& d_{17}\left(1-\frac{\bc}{f}+\frac{\bc^2}{f^2}+\ldots\right)
& d_{23}\FF_{23}(\chi)
& d_{23}
\\
d_{13}\FF_{13}(\chi)
& d_{13}
& d_{18}\FF_{18}(\chi)
& d_{18}\left(1-\frac{\bc}{f}+\frac{\bc^2}{f^2}+\ldots\right)
& d_{24}\FF_{24}(\chi)
& d_{24}
\\
d_{14}\FF_{14}(\chi)
& d_{14}
& d_{19}\FF_{19}(\chi)
& d_{19}\left(1-\frac{\bc}{f}+\frac{\bc^2}{f^2}+\ldots\right)
& d_{25}\FF_{25}(\chi)
& d_{25}\left(1-\frac{\bc}{f}+\frac{\bc^2}{f^2}+\ldots\right)
\\
d_{15}\FF_{15}(\chi)
& d_{15}
& d_{21}\FF_{21}(\chi)
& d_{21}\left(1-\frac{2\bc}{f}+\frac{3\bc^2}{f^2}+\ldots\right)
& d_{26}\FF_{26}(\chi)
& d_{26}
\\
d_{16}\FF_{16}(\chi)
& d_{16}
& d_{22}\FF_{22}(\chi)
& d_{22}\left(1-\frac{2\bc}{f}+\frac{3\bc^2}{f^2}+\ldots\right)
& 
& 
\\\hline
\end{tabular}}
\caption{
\it Explicit expressions for the products $d_{i}\FF_{i}(h)$ of the low-energy custodial breaking operators.
}
\label{tableComparisonsCustViol}
\end{table*}

The custodial breaking couplings will be reported in Tab.~\ref{tableComparisonsCustViol} only for the DEFT basis. Indeed, the minimal $SO(5)/SO(4)$ CH model does not allow these couplings to appear, and for the SMEFT basis only the operator $\cO_{\Phi,1}$ is custodial breaking: projecting it into the HEFT basis, one finds that the function $\cF_T(h)$ receives the following contribution,
\be
c_T\cF^\text{SMEFT}_T(h)=-\dfrac{(4\pi)^2}{4}\dfrac{v^2}{\Lambda^2}c_T\left(1+\dfrac{h}{v}\right)^2\,.
\ee
This can be compared with the corresponding contribution in the DEFT basis:
\be
d_T\FF^\text{DEFT}_T(\chi)=d_T\left(1+\dfrac{\bc}{f}\right)^2\,.
\label{FTDilaton}
\ee
In the specific case in which the conformal sector does preserve custodial symmetry, then all the custodial breaking operators are expected to be further suppressed and its dedicated analysis turns out to be less interesting. 

Finally, self-couplings of the dilaton are expected in general to be different from the SMEFT and HEFT cases. However, without entering into details of specific realisations, both for the dilaton and the composite Higgs, any dedicated analysis would not be conclusive. For a discussion in this respect see, for example, Ref.~\cite{Vecchi:2010gj}.

\subsection{Phenomenology Avenues}

As previously discussed in Eqs.~(\ref{FCDilaton}), (\ref{FYDilaton}) and (\ref{FTDilaton}) for the lowest order Lagrangian, the Higgs-like dilaton has couplings very similar to the ones of the SM Higgs, at least for the shape of these couplings: the bounds on the $v/f$ factor present in these couplings from data on Higgs interactions with longitudinal components of gauge bosons and with fermions indicate that the scale of the scale symmetry breaking $f$ must be close to the EWSB one $v$. Besides this, the numerical value of the correlation between 0, 1, and 2 dilaton insertions in a specific coupling is the same as in the SM. This represents a net prediction of the DEFT Lagrangian with respect to the HEFT and SMEFT cases: the ratio of the branching ratios of the dilaton that decay into two gauge bosons and into two fermions is expected to be 1. Moreover, it opens the possibility to tell the dilaton from other possibilities by investigating the growth with the energy of amplitudes involving two longitudinal components of the SM gauge boson. Indeed, in the dilaton case as in the SM, the amplitudes for $VV\to VV$ and $VV\to \bc\bc$ among others, with $V$ a massive gauge boson, do not grow as $E^2$, where $E$ represents the energy of the subprocess. The opposite occurs for the equivalent processes in the SMEFT and HEFT cases. See Refs.~\cite{Ballestrero:2009vw,Vecchi:2010gj} for a dedicated analysis in this respect. 

Going beyond the lowest order Lagrangian, Tabs.~\ref{tableComparisonsCustCons} and \ref{tableComparisonsCustViol} allow to discuss signals that could be able to disentangle a Higgs-like dilaton from other possibilities. The features that jump to the eyes are:
\begin{itemize} 
\item[i)] the specific linear combinations of $h$ and $\varphi$ in the $\cF(h)$ functions and of $\bc$ in the $\FF_i(\chi)$ functions;
\item[ii)] the presence of a few pure-gauge couplings that do not come along with $\bc$ insertions, while these same couplings appear with $h$ insertions in the SMEFT and in the CH setup: $\PP_2$, $\PP_3$, $\PP_6$, $\PP_9$, $\PP_{11-16}$, $\PP_{23}$, $\PP_{24}$ and $\PP_{26}$;
\item[iii)] the presence of $\bc$ insertions couplings, that are instead predicted to be only pure-gauge in the SMEFT and in the CH framework: $\PP_{WWW}$, $\PP_{GGG}$, $\PP_{DB}$, $\PP_{DW}$ and $\PP_{DG}$;
\item[iv)] the independence of the couplings described in each line of the tables due to the $d_i$ coefficients, while SMEFT and the CH Lagrangian predict correlations between couplings of different lines: for example, $\cF_2(h)$ and $\cF_4(h)$, or $\cF_3(h)$ and $\cF_5(h)$, etc.. 
\item[v)] all the custodial breaking terms in Tab.~\ref{tableComparisonsCustViol} are not expected either in the $\dim=6$ SMEFT Lagrangian, nor in the minimal $SO(5)/SO(4)$ CH model.
\end{itemize}

Concerning point i), the specific correlations between the 0, 1, 2, $\ldots$, dilaton insertion in a given coupling are distinct from those in the SMEFT and in CH models: this could indeed discriminate between the different frameworks, once the sensitivity on Higgs couplings will increase. Several studies on phenomenological signals of a Higgs-like dilaton at colliders have been presented. Of particular interest is the process $\bc\bc\to \bc\bc$ for the DEFT case to be compared with its sibling $hh\to hh$ in the SMEFT and HEFT cases: in the DEFT and HEFT cases the amplitude grows with energy as $E^4$ (due to the operators $\PP_{D\chi}$ and $\cP_{DH}$, respectively), while only as $E^2$ in the SMEFT (due to the operator $\cO_{\Phi,2}$). See, for example, Refs.~\cite{Goldberger:2008zz,Bellazzini:2012vz} for dedicated analyses on these observables.

The absence of dilaton couplings reported in point ii) is connected to another type of potentially interesting phenomenological signals. For example, the operator $\PP_2$ describes only $\gamma-W-W$ and $Z-W-W$ pure-gauge couplings with specific Lorentz contractions, and in particular no dilaton insertion is expected in the DEFT Lagrangian; in the SMEFT and CH setups, these pure-gauge couplings arise due to the $\cP_2$ operator and are predicted to be correlated to the Higgs couplings $\gamma-W-W-h$ and $Z-W-W-h$, respectively. In a one-operator-at-a-time analysis, the absence of the dilaton insertion is a smoking gun for this scenario; on the other side, in a multi-operatorial analysis, the discriminating power of this feature is much reduced and the impact of this operator should be analysed together with the effects of the other Lagrangian operators.

The case described in point iii) is the opposite with respect to what was just discussed about point ii): couplings with dilaton insertions, whose corresponding couplings with $h$ are not expected in the  SMEFT and CH setup, at the expanding order considered. The discussion and conclusions of the previous paragraph also apply to this case, after the necessary rephrasing.

Point iv) states that several couplings that are expected to be correlated (at a given uncertainty) in the SMEFT and CH framework, are instead decorrelated in the DEFT case. This is the case, for example, for $\cP_2$ and $\cP_4$, or $\cP_3$ and $\cP_5$. The kind of decorrelation of the DEFT is the same as that of the generic HEFT Lagrangian discussed in Refs.~\cite{Brivio:2013pma,Brivio:2016fzo}: see in particular Fig.~3 of Ref.~\cite{Brivio:2016fzo}, which also applies to the DEFT Lagrangian, after redefining the $\Sigma$ and $\Delta$ functions:
\be
\begin{aligned}
\Sigma_B\equiv \dfrac{1}{\pi g t_\theta}(2d_2-d_4)\,,\quad
\Sigma_W\equiv \dfrac{1}{2\pi g}(2d_3+d_5)\,,\\
\Delta_B\equiv \dfrac{1}{\pi g t_\theta}(2d_2+d_4)\,,\quad
\Delta_W\equiv \dfrac{1}{2\pi g}(2d_3-d_5)\,,
\end{aligned}
\ee
where $t_\theta$ is the tangent of the Weinberg angle. The two $\Delta$'s parameters are zero in the $\dim=6$ SMEFT Lagrangian in consequence of gauge invariance and the doublet nature of the Higgs. $\Delta_B=0=\Delta_W$ also in the minimal $SO(5)/SO(4)$ CH model once $v^4/f^4$ terms are neglected, due to the almost exactly doublet embedding of the Higgs. Then, deviations from zero larger than $v^4/f^4$ cannot be explained with $\dim=8$ SMEFT contributions or with contributions from the $SO(5)/SO(4)$ CH Lagrangian, while they could be compatible with the DEFT description. The orthogonal combinations, $\Sigma_B$ and $\Sigma_W$ will instead represent deviations from the exact SM case and cannot distinguish among $\dim=6$ SMEFT, $SO(5)/SO(4)$ CH or DEFT contributions. Further details can be found in Ref.~\cite{Brivio:2016fzo}.

Finally, the custodial breaking contributions in point v) are only present in the DEFT Lagrangian, while they are not expected either in the $\dim=6$ SMEFT Lagrangian, ot in the minimal $SO(5)/SO(4)$ CH model. They represent interesting possibilities to tell the dilaton from the other frameworks: this is the case of the pure-gauge operator $\cP_{14}$, equal to $\PP_{14}$, discussed in Ref.~\cite{Brivio:2013pma};  this operator contributes to the anomalous triple gauge vertex
\be
i g_Z^5 \epsilon^{\mu\nu\rho\lambda}\left(W_\mu^+\derp_\rho W_\nu^--W_\nu^-\derp_\rho W_\mu^+\right)Z_\sigma\,,
\ee
that originates $WW$ and $WZ$ pair production. The second observable has been studied in Ref.~\cite{Brivio:2013pma} by analysing the reaction
\be
pp\to\ell^{\prime\pm}\ell^+\ell^-\,E_T^\text{miss}\,,
\ee
where $\ell^{(\prime)}=e$ or $\mu$. The predicted values of the cross sections for this process considering $7$ TeV, $8$ TeV, and $14$ TeV center of mass energy have been computed and the expected number of events over the background has been illustrated in Fig.~3 of Ref.~\cite{Brivio:2013pma}. The results are that the precision on $g_Z^5$ at the LHC 7 and 8 TeV runs is already higher than the present direct bounds stemming from LEP, and therefore LHC already has the potential to discriminate a Higgs-like dilaton from an (almost) exactly EW doublet Higgs. 

%
%
\section{Conclusions}
\label{Sect:Conclusions}

The CP-even complete bosonic effective Lagrangian for a Higgs-like dilaton has been constructed considering an expansion up to the first subleading order. This basis has been compared with the $\dim=6$ SMEFT Lagrangian and with the effective Lagrangian for the minimal $SO(5)/SO(4)$ CH model. 

Five distinct features that could distinguish among the different Higgs descriptions have been discussed: they are related to the presence (or absence) of specific correlations among dilaton insertions, which are distinct from the correlations among $h$ insertions in the SMEFT and CH setups; moreover, they are due to the larger number of independent operators in the DEFT basis. Finally, they follow the fact that custodial breaking contributions are only predicted in the DEFT Lagrangian, at the considered expansion order. 

Previous studies on the generic HEFT Lagrangian also hold for the DEFT Lagrangian: correlations/decorrelations between pure-gauge couplings and dilaton-gauge bosons interactions, and anomalous triple gauge boson couplings only predicted to be relevant in the DEFT case. They represent promising signals to disentangle a Higgs-like dilaton from an (almost) exact EW doublet Higgs at colliders.


\acknowledgments

We thank Rodrigo Alonso, Juan Gonz\'alez-Fraile and Concha Gonz\'alez-Garc\'ia for very useful discussions.\\

L.M. acknowledges partial financial support by the European Union through the FP7 ITN INVISIBLES (PITN-GA-2011-289442), by the Horizon2020 RISE InvisiblesPlus 690575, by CiCYT through the projects FPA2012-31880 and FPA2016-78645, and by the Spanish MINECO through the Centro de excelencia Severo Ochoa Program under grant SEV-2012-0249. The work of L.M. is supported by the Spanish MINECO through the ``Ram\'on y Cajal'' programme (RYC-2015-17173).

\appendix

%
%
\section{The HEFT and DEFT Building Rules}
\label{APP:Counting}

This section describes the rules that have been adopted for the construction of the HEFT and DEFT Lagrangians. 

Focussing first on the HEFT, the division in Eq.~(\ref{HEFTLAG}) and the specific choice of the operators entering $\LL_{h}^{(0)}$ and $\Delta \LL_{h}$ follow a specific set of rules that have been described at length in Ref.~\cite{Brivio:2016fzo}. Different counting rules would have eventually led to a different ordering of some operators, and therefore a deep understanding of the HEFT is mandatory. The part that follows is a resume of the discussion presented in Ref.~\cite{Brivio:2016fzo}.

The HEFT can be seen as a fusion of the traditional linear description -- based on canonical mass dimensions -- for the transverse components of the gauge bosons and for fermions, and of the chiral perturbation approach -- based on counting derivatives -- associated to the SM GBs and the physical $h$. Indeed, the latter enters the Lagrangian via the adimensional functions $\cF(h)$, that play the same role as the adimensional GB matrix field $\U(x)$: in concrete CH models, the $\cF(h)$ are trigonometric functions. $\U(x)$ and $\cF(h)$ produce operators with different canonical mass dimensions and contribute to physical observables, such as cross sections, at different orders in energy$/\Lambda$~\cite{Gavela:2016bzc}: $\U(x)$ and $\cF(h)$ cannot be treated as having a homogeneous mass dimension for power counting purposes.

As the HEFT is a merging between linear and chiral descriptions, the counting rules which would apply singularly to each of the expansions hold simultaneously for the HEFT~\cite{Gavela:2016bzc}. The consequences can be seen even at the level of $\LL_{h}^{(0)}$ in Eq.~(\ref{Lag0}): it does not strictly respect the chiral expansion as it contains both operators with two derivatives, such as $\Tr(\V_\mu \V^\mu)\cF_C(h)$, and the gauge boson kinetic terms; on the other hand, $\LL_{h}^{(0)}$ does not follow an expansion in canonical dimensions, as an infinite series of $h/f$ is encoded into the functions $\cF_i(h)$.

The renormalisability conditions that enter in the determination of $\Delta \LL_{h}$ are also different. In the linear
expansion, the divergences generated by an $n$-loop diagram containing one single $d = 6$ vertex can be reabsorbed by other $d = 6$ operators and, in particular, do not require the introduction of any higher-dimensional operator. In the chiral case, instead, 1-loop diagrams with $n$ insertions of a two-derivative coupling, typically listed in the LO Lagrangian,
produce divergences that force the introduction of operators with four-derivatives, which generically constitute the NLO Lagrangian.

Last but not least, the HEFT presents several scales in the expansion. Besides the cut-off of the theory $\Lambda$, one can identify the presence of the SM GB scale $f_\pi$ and of the $h$-scale $f$, which are typically distinct from each other and correspond to different physical quantities. Moreover, the HEFT is affected by the well-known fine-tuning associated to the EW scale $v$: for example, the parameters $a_C$ and $b_C$ in Eq.~(\ref{FC}) are expected to be arbitrary numbers of order 1, while phenomenologically they are expected to be close to 1; similarly for the Yukawa couplings, the HEFT allows fermion-Higgs couplings to be independent from fermion masses, while data suggest and an alignment among these classes of couplings.

In conclusion, the HEFT expansion depends on more than one single parameter, and depends on the typical energy scale associated to the observables considered in the phenomenological analysis~\cite{Gavela:2016bzc,Brivio:2016fzo}. 
Moreover, the HEFT building blocks have completely different properties and therefore cannot be treated homogeneously in full generality. For this reason, the division between $\LL_{h}^{(0)}$ and $\Delta \LL_{h}$ and the choice of the NLO Lagrangian operators have not been based on a single-parameter counting rule, whose applicability is not valid in full generality and that would necessarily lead to {\it ad hoc} assumptions. Instead, the selection performed in Ref.~\cite{Brivio:2016fzo}, is based on NDA~\cite{Manohar:1983md,Cohen:1997rt,Gavela:2016bzc}: this power counting rule accounts for both the mass dimensions of the single building blocks, and the type and number of fields entering in an operator. It reduces to the following NDA master formula, first proposed in Ref.~\cite{Manohar:1983md} and later modified in Refs.~\cite{Cohen:1997rt} and \cite{Gavela:2016bzc}: following the notation of Ref.~\cite{Gavela:2016bzc},
\be
\frac{\Lambda^4}{16 \pi^2 } \left[\frac{\partial}{\Lambda}\right]^{N_p}  \left[\frac{ 4 \pi\,  \phi}{ \Lambda} \right]^{N_\phi}
 \left[\frac{ 4 \pi\,  A}{ \Lambda } \right]^{N_A}  \left[\frac{ 4 \pi \,  \psi}{\Lambda^{3/2}}\right]^{N_\psi} \left[ \frac{g}{4 \pi }  \right]^{N_g}\,,
\label{NDAMasterFormula}
\ee
where $\phi$ represents either the SM GBs or $h$, $\psi$ a generic fermion, $A$ a generic gauge field, and $g$ the gauge couplings. $N_i$ counts the number of fields of each type entering a given operator.

$\LL_{h}^{(0)}$ contains all the operators that result unsuppressed according to the NDA master formula. These are the kinetic terms for gauge bosons, SM GBs, $h$, and fermions, plus the Yukawa interactions, the $h$ scalar potential and the QCD theta term (the latter has been written suppressed by $g_s^2/(4\pi)^2$ following the traditional notation, but no such suppression is required following Eq.~(\ref{NDAMasterFormula})). Although any of such operators could appear multiplied by a function $\cF(h)$, this is not the case for the kinetic terms of fermions and of the physical Higgs, as their $\cF(h)$ can be reabsorbed inside the generic functions $\cF_C(h)$ and $\cY_{Q,L}(h)$. The kinetic terms of the gauge bosons are not written with an associated $\cF(h)$ assuming that the transverse components of the gauge bosons do not couple strongly to the EWSB sector~\cite{Contino:2010mh}: this is a phenomenological assumption that does not follow from the NDA counting rule, but reflects the explicit choice of considering the HEFT Lagrangian as the low-energy description of theories where the gauge bosons have a distinct origin in the EWSB sector. 

$\LL_{h}^{(0)}$ does not contain the operator $\cP_T$, in Eq.~(\ref{P_T}), although it comes unsuppressed and therefore should have been treated in the same way as $\tr(\V_\mu\V^\mu)$. This is simply a stylistic choice that has been adopted in Ref.~\cite{Brivio:2016fzo}: if $\cP_T$ was introduced in $\LL_{h}^{(0)}$, its coefficient would have been suppressed at the level of the percentfrom the constraints on the EW precision parameter $T$; its effects then would have been completely negligible in the rest of the Lagrangian as any contribution in the equations of motion and in loop-induced observables would have come weighted by powers of its coefficient. Instead, inserting $\cP_T$ in $\Delta \LL_{h}$ simplifies the definition of the equations of motion and the computations of loop-contributions, without changing any aspect of the phenomenological analysis. On the other hand, $\cP_T$ needs to belong to $\Delta \LL_{h}$: it has been shown long time ago in Refs.~\cite{Longhitano:1980iz,Longhitano:1980tm} considering the EW chiral Lagrangian, which is the effective Lagrangian constructed with the SM spectrum with the exception of the physical $h$, that $\cP_T$ arises as a counter term necessary to absorb the 1-loop divergences originated by the renormalisation of a lowest order Lagrangian defined as in Eq.~(\ref{Lag0}) switching off fermions and $h$.

$\LL_{h}^{(0)}$ does not contain the other two operators that come unsuppressed, $\cP_1$ and $\cP_{12}$. Inserting these operators into $\LL_{h}^{(0)}$ would have modified the kinetic terms of the gauge bosons, requiring a field redefinition in order to obtain canonical kinetic terms. After this redefinition, $\cP_1$ and $\cP_{12}$ disappear from $\LL_{h}^{(0)}$, but need to be inserted into $\Delta \LL_{h}$ in order to absorb 1-loop divergences originated during the renormalisation of $\LL_{h}^{(0)}$.

In summary, $\LL_{h}^{(0)}$ contains all the operators that are not suppressed by any scale according to NDA. The exceptions are well justified in order to guarantee that $\LL_{h}^{(0)}$ contains canonical kinetic terms and to simplify the equations of motion and the loop computations.

The second part of the HEFT Lagrangian $\Delta \LL_{h}$ contains the couplings excluded from $\LL_{h}^{(0)}$ and all the structures necessary to absorbe the 1-loop divergences that originated in the renormalisation of $\LL_{h}^{(0)}$ (the list of custodial preserving operators have been studied in Refs.~\cite{Espriu:2013fia,Delgado:2013loa,Delgado:2014jda,Gavela:2014uta,Guo:2015isa,Alonso:2015fsp}, while for the custodial breaking ones it is necessary to generalise from the results in Refs.~\cite{Longhitano:1980iz,Longhitano:1980tm}). The first ones are: $\cP_T$, $\cP_1$, and $\cP_{12}$, discussed previously in this appendix; $\cP_{B}$, $\cP_{W}$, and $\cP_{G}$, that describe couplings between the physical $h$ and the transverse components of the SM gauge bosons, which could be originated at 1-loop even if the assumption of the absence of any tree level coupling of this kind; $\cP_{h}$ that represents a modification of the $h$ kinetic term and that can be reabsorbed by an $h$ redefinition. As already mentioned, these operators are not suppressed by any scale according to the NDA, but have been left out from $\LL_{h}^{(0)}$, as motivated above. 

The other operators in $\Delta \LL_{h}$ are suppressed either by $4\pi$ or $(4\pi)^2$, or by $\Lambda^2$. To see explicitly that this rule is respected without any exception, it is however necessary to reconstruct the $\cF(h)$ function in all the couplings, by substituting $\Lambda=4\pi f$ and collecting powers of $h/f$; this is consistent with the fact that $\U(x)$ is kept in the compact notation instead of having been written explicitly in terms of the GB fields. As an example, a few operators have been reported by rewriting the explicit $h$ dependence in terms of the adimensional fraction $h/f$ and extracting the overall suppression:
\be
\begin{aligned}
\cP_{\square H}&\to \dfrac{1}{(4\pi)^2}\left(\square\dfrac{h}{f}\right)^2\\
\cP_{\Delta H}&\to\frac{1}{(4\pi)^2}\left(\derp_\mu \dfrac{h}{f}\right)\left(\derp^\mu \dfrac{h}{f}\right)\left(\Box \dfrac{h}{f}\right)\\
\cP_{DH}&\to\frac{1}{(4\pi)^2} \left(\left(\derp_\mu \dfrac{h}{f}\right)\left(\derp^\mu \dfrac{h}{f}\right)\right)^2\\
\cP_{4}  &\to \dfrac{i}{4\pi}B_{\mu\nu}\Tr(\T\V^\mu)\,\derp^\nu \dfrac{h}{f}\\
\cP_{7} &\to\dfrac{1}{(4\pi)^2}\Tr\left(\V_\mu\,\V^\mu\right)\derp_\nu\derp^\nu \dfrac{h}{f}\\
\cP_{8} &\to\dfrac{1}{(4\pi)^2}\Tr\left(\V_\mu\,\V_\nu\right)\derp^\mu \dfrac{h}{f}\,\derp^\nu \dfrac{h}{f}\,,
\end{aligned}
\ee
and similarly for the other operators in Eq.~(\ref{HEFTOperators}). Structures with stronger suppressions than $1/(4\pi^2)$ or $\Lambda^2$ according to the NDA would be listed in a higher order part of the full Lagrangian. 

The suppression of the operators in $\Delta \LL_{h}$ reflects from one side the renormalisation of the chiral sector, according to the chiral perturbation theory, and from the other the possible presence of new physics contributions, which may be due to the tree level exchange of new particles. Once more, this makes it evident that HEFT is a marriage between two approaches, the linear and the chiral ones, and that a simplistic approach to encode the expansion within a single parameter counting is not natural, while the relevant issue is the definition of a reasonable starting Lagrangian and the identification of the structures necessary to absorb the divergences that originated at 1-loop. 

An interesting aspect, that is manifest when fermions are also considered (see Ref.~\cite{Brivio:2016fzo}), is that not all the operators with the same suppression are necessary for removing 1-loop divergences: an example of this is a set of four-fermions couplings or the dipole operators. For this reason, it is customary to insert into the higher order parts of the Lagrangian all the operators necessary to reabsorb the divergences that originate in the renormalisation of the lowest parts of the Lagrangian, together with operators with similar suppressions.\\

The DEFT Lagrangian is constructed following exactly the same reasoning as that of the HEFT construction described in detail in this appendix. The division in Eq.~(\ref{DEFTLAG}) and the selection of operators entering $ \Delta \LL_{\chi}$ automatically follow by adopting the NDA counting and following the same rules as for the HEFT, except for the fact that $\chi$ must guarantee the scale invariance.

%
%
\section{The SMEFT Lagrangian}
\label{APP:SMEFT}

The SMEFT Lagrangian~\cite{Buchmuller:1985jz,Grzadkowski:2010es} is constructed by means of the SM particles and in particular with the Higgs field belonging to a doublet of the $SU(2)_L$ symmetry, Eq.~(\ref{HiggsEq}). When restricting only on the bosonic sector and only on operators of $\dim\leq6$, the list of terms defining a non-redundant basis counts 11 elements describing Higgs interactions: the so-called Hagiwara-Ishihara-Szalapski-Zeppenfeld (HISZ) basis~\cite{Hagiwara:1993ck,Hagiwara:1996kf}, plus an additional 5 elements containing pure-gauge couplings~\cite{DeRujula:1991ufe}. Considering these restrictions, the Lagrangian can be written as a sum of two terms,
\be
\LL_\text{SMEFT} = \LL_{SM}+\Delta\LL_\text{SMEFT}\,,
\label{Llinear}
\ee
where $\LL_{SM}$ is the SM Lagrangian and
\be
\Delta \LL_\text{SMEFT} = \sum_i \frac{c_i}{\Lambda^2} \cO_i\,, 
\label{DeltaLlinear}
\ee
with $c_i$ being order one parameters and $\cO_i$ defined as follows:
\be
\begin{aligned}
\cO_{BB} &= (4\pi)^2\Phi^{\dagger} B_{\mu \nu} B^{\mu \nu} \Phi\\\
\cO_{WW} &= (4\pi)^2\Phi^{\dagger} \W_{\mu \nu} \W^{\mu \nu} \Phi \\
\cO_{GG} &=(4\pi)^2\Phi^\dagger \Phi \,G^\alpha_{\mu\nu} G^{\alpha\,\mu\nu}\\
\cO_{BW} &=  (4\pi)^2\Phi^{\dagger} B_{\mu \nu} \W^{\mu \nu} \Phi \\
\cO_W  &= 4\pi i(\DL_{\mu} \Phi)^{\dagger} \W^{\mu \nu}  (\DL_{\nu} \Phi)\\ 
\cO_B  &= 4\pi i(\DL_{\mu} \Phi)^{\dagger} B^{\mu \nu}  (\DL_{\nu} \Phi)\\
\cO_{\Phi,1} &= (4\pi)^2 \left (\DL_\mu \Phi \right)^\dagger \Phi\  \Phi^\dagger \left (\DL^\mu \Phi \right )\\
\cO_{\Phi,2} &=(4\pi)^2 \frac{1}{2} \partial^\mu\left ( \Phi^\dagger \Phi \right)
\partial_\mu\left ( \Phi^\dagger \Phi \right)\\
\cO_{\Phi,3}&=(4\pi)^4\left(\Phi^\dag\Phi\right)^3\\
\cO_{\Phi,4} &= (4\pi)^2\left (\DL_\mu \Phi \right)^\dagger \left(\DL^\mu\Phi\right)\left(\Phi^\dagger\Phi \right )\\
\cO_{\Box \Phi}&=\left(\DL_\mu \DL^\mu\Phi\right)^\dag\left(\DL_\nu \DL^\nu\Phi\right)\\
\cO_{3W} &=4\pi\varepsilon_{abc}W_\mu^{a\nu}W_\nu^{b\rho} W_{\rho}^{c\mu}\\
\cO_{3G} &=4\pi f_{\alpha\beta\gamma}{\mathcal G}_\mu^{\alpha\nu}{\mathcal G}_\nu^{\beta\rho} {\mathcal G}_{\rho}^{\gamma\mu}\\
\cP_{DB} & = \left(\derp^\mu B_{\mu\nu}\right)\left(\derp_\rho B^{\rho\nu}\right)\\
\cP_{DW} & = \left(\cD^\mu W_{\mu\nu}\right)^a\left(\cD_\rho W^{\rho\nu}\right)^a\\
\cP_{DG} & = \left(\cD^\mu G_{\mu\nu}\right)^\alpha\left(\cD_\rho G^{\rho\nu}\right)^\alpha\,.
\end{aligned}
\label{LinearOpsGauge} 
\ee
In the previous expressions, $\DL_\mu\Phi\equiv \left(\partial_\mu+ \frac{i}{2} g' B_\mu + \frac{i}{2}g\sigma_i W^i_\mu \right)\Phi $ is the covariant derivative applied to the Higgs doublet.  Among these operators, only $\cO_{\Phi,1}$ is custodial breaking, while the rest are custodial preserving. All the operators are written according to the Naive Dimensional Analysis normalisation~\cite{Gavela:2016bzc}.

The operators $\cO_{\Phi,1}$, $\cO_{\Phi,2}$, and $\cO_{\Phi,4}$ contribute to the Higgs kinetic term:
\be
\dfrac{1}{2}\derp_\mu h\derp^\mu h\left(1+\dfrac{(4\pi)^2}{2}\dfrac{(v+h)^2}{\Lambda^2}\left(c_{\Phi1}+2c_{\Phi2}+c_{\Phi4}\right)\right)\,.
\ee 
It is then necessary to perform a field redefinition to move to the basis of canonical kinetic terms (see App.~B in Ref.~\cite{Brivio:2016fzo}): neglecting terms proportional to products of two or more $c_i$, the transformation reduces to
\be
h\to h \left(1-\frac{(4\pi)^2}{12}\left(c_{\Phi1}+2c_{\Phi2}+c_{\Phi4}\right)\dfrac{3v^2+3vh+h^2}{\Lambda^2}\right)\,.
\ee
A priori, all the Higgs couplings in the full Lagrangian would be affected by this transformation; however, neglecting all the terms proportional to products of two or more $c_i$, only the Higgs interactions present in the SM Lagrangian will be modified, i.e. the couplings with two longitudinal components of the EW gauge bosons and the Yukawa terms.

%
%
\section{The minimal $SO(5)/SO(4)$ CH Lagrangian}
\label{APP:CH}

The Lagrangian for the $SO(5)/SO(4)$ minimal CH model has been known already for some time~\cite{Agashe:2004rs,Alonso:2014wta,Hierro:2015nna,Feruglio:2016zvt,Gavela:2016vte}, and the notation used in Refs.~\cite{Alonso:2014wta,Hierro:2015nna} will be adopted in what follows. Mimicking the Appelquist-Longhitano-Feruglio convention, one can define a field matrix $\Omega(x)$ containing all the GBs that originate from the $SO(5)/SO(4)$ spontaneous breaking and transforming under the global groups $SO(5)$ and $SO(4)$ as
\be
\Omega(x)\to\gtt\,\Omega(x)\, \htt^{-1}\,,
\ee
where $\gtt$ is an element of $SO(5)$ and $\htt$ is an element of $SO(4)$. It is then possible to define a ``squared'' non-linear field matrix $\SH(x)$ transforming only under $SO(5)$:
\be
\SH(x)\equiv \Omega(x)^2\,,\qquad
\SH(x)\to \gtt\SH(x)\gtt_R^{-1}\,,
\ee
where $\gtt_R$ is the grading of $\gtt$ (see Refs.~\cite{Alonso:2014wta,Hierro:2015nna} for further details). It is now possible to define the quantity $\Vt$, which is equivalent to the vector chiral field $\V$ of the EW chiral Lagrangian:
\be
\Vt_\mu=\left(\D_\mu\SH\right)\SH^{-1}\,,
\qquad
\Vt_\mu\to\gtt\Vt_\mu\gtt^{-1}\,,
\ee
where $\D_\mu$ is the covariant derivative acting on $\SH$.

Next, one can consider the gauging of the SM group, $SU(2)_L\times U(1)_Y\subset SO(5)$ and $SU(2)_L\times U(1)_Y\not\subset SO(4)$, and the embedding of the SM gauge vector boson into the $SO(5)$ representations, 
\be
\WLt_\mu\equiv W^a_\mu \,Q^a_L \qquad {\rm and} \qquad \BLt_\mu \equiv B_\mu \,Q_Y\,,
\label{WBSU(5)}
\ee
where $Q^a_L$ and $Q_Y$  denote the embedding in $SO(5)$ of the $SU(2)_L\times U(1)_Y$ generators.  

Finally, the CP-even EW high-energy chiral Lagrangian describing bosonic interactions, $\LL_{CH}$, can be written as
\be
\LL_{CH}=\LL^{0}_{CH}+\Delta\LL_{CH}\,,
\label{LLG}
\ee
where
\begin{align}
\LL^{0}_{CH}=&-\frac{1}{4} \Tr\left(\BLt_{\mu \nu}\BLt^{\mu \nu}\right)-\frac{1}{4} \Tr\left(\WLt_{\mu \nu}\WLt^{\mu \nu}\right)+\nn\\
&-\dfrac{1}{4} \GGd^\alpha \mathcal{G}^{\alpha\,\mu\nu}-\frac{f^2}{4}\Tr\left(\Vt_\mu\Vt^\mu\right)+i\bar{Q}_L\slashed{D}Q_L+\nn\\
&+i\bar{Q}_R\slashed{D}Q_R+i\bar{L}_L\slashed{D}L_L
+i\bar{L}_R\slashed{D}L_R+\\
&-\dfrac{f}{\sqrt2}\left(\bar{Q}_L\U Y^{(0)}_Q Q_R\sin\frac{\varphi}{f}+\hc\right)+\nn\\
&-\dfrac{f}{\sqrt2}\left(\bar{L}_L\U Y^{(0)}_L L_R\sin\frac{\varphi}{f}+\hc\right)\,,\nn
\end{align}
and 
\be
\Delta\LL_{CH}=\ct_{B\Sigma}\cAt_{B\Sigma}+\ct_{W\Sigma}\cAt_{W\Sigma}+\sum_{i=1}^6\ct_i\,\cAt_i\,.
\ee
The operators appearing in the previous expression are defined by
\pagebreak
\be
\begin{aligned}
\cAt_{B\Sigma} &=\Tr\left(\SH\BLt_{\mu \nu}\SH^{-1}\BLt^{\mu \nu}\right)\\
\cAt_{W\Sigma} &=\Tr\left(\SH\WLt_{\mu \nu}\SH^{-1}\WLt^{\mu \nu}\right)\\
\cAt_{1}  &= \Tr\left(\SH\BLt_{\mu \nu}\SH^{-1}\WLt^{\mu \nu}\right)\\
\cAt_{2}  &=\dfrac{i}{4\pi}\Tr\left(\BLt_{\mu \nu}\left[\Vt^\mu,\Vt^\nu\right]\right)\\
\cAt_{3} &=\dfrac{i}{4\pi}\Tr\left(\WLt_{\mu\nu}\left[\Vt^\mu,\Vt^\nu\right]\right)\\
\cAt_{4} &= \dfrac{1}{(4\pi)^2}\left(\Tr\left(\Vt_\mu\,\Vt^\mu\right)\right)^2\\
\cAt_{5} &= \dfrac{1}{(4\pi)^2}\left(\Tr\left(\Vt_\mu\,\Vt_\nu\right)\right)^2\\
\cAt_{6} &= \dfrac{1}{(4\pi)^2}\Tr\left((\cD_\mu\Vt^\mu)^2 \right)\,,
\end{aligned}
\label{AppelquistBasisG}
\ee
with the EW covariant derivative applied to $\Vt$ as
\be
\cD_\mu\Vt^\mu= \derp_\mu \Vt^\mu+i\, g\left[\WLt_\mu,\Vt^\mu\right]+i\, g'\left[\BLt_\mu,\Vt^\mu\right]\,.
\ee
The coefficients $\ct_i$ are expected to be all of the same order of magnitude, according to the effective field theory approach, except for the coefficients of the operators in $\LL^{(0)}_{CH}$ which are taken equal to $1$, leading to canonical kinetic terms.  The expansion is stopped at the first order in inverse powers of the scale associated to the $SO(5)/SO(4)$ breaking, and therefore no $6$-derivative operators have been considered in $\Delta \LL_{CH}$.

Couplings to gluons are expected to appear due to fermionic loops: however, we are not aware of any explicit computation of such contributions present in the literature and it is beyond the scope of the present paper to perform such a computation.



\begin{thebibliography}{10}

\bibitem{Aad:2012tfa}
{\bf ATLAS} Collaboration, G.~Aad {\em et.~al.},  Phys. Lett. {\bf B716} (2012)
  1--29, [\href{http://xxx.lanl.gov/abs/1207.7214}{{\tt arXiv:1207.7214}}].

\bibitem{Chatrchyan:2012xdj}
{\bf CMS} Collaboration, S.~Chatrchyan {\em et.~al.},  Phys. Lett. {\bf B716}
  (2012) 30--61, [\href{http://xxx.lanl.gov/abs/1207.7235}{{\tt
  arXiv:1207.7235}}].

\bibitem{Mariotti:2016owy}
C.~Mariotti and G.~Passarino,  \href{http://xxx.lanl.gov/abs/1612.00269}{{\tt
  arXiv:1612.00269}}.
  
\bibitem{Weinberg:1978kz}
S.~Weinberg, Physica A {\bf 96} (1979) 327.  
  
\bibitem{Buchmuller:1985jz}
W.~Buchmuller and D.~Wyler,  Nucl. Phys. {\bf B268} (1986) 621--653.

\bibitem{Grzadkowski:2010es}
B.~Grzadkowski, M.~Iskrzynski, M.~Misiak, and J.~Rosiek,  JHEP {\bf 10} (2010)
  085, [\href{http://xxx.lanl.gov/abs/1008.4884}{{\tt arXiv:1008.4884}}].

\bibitem{Feruglio:1992wf}
F.~Feruglio,  Int. J. Mod. Phys. {\bf A8} (1993) 4937--4972,
  [\href{http://xxx.lanl.gov/abs/hep-ph/9301281}{{\tt hep-ph/9301281}}].

\bibitem{Contino:2010mh}
R.~Contino, C.~Grojean, M.~Moretti, F.~Piccinini, and R.~Rattazzi,  JHEP {\bf
  05} (2010) 089, [\href{http://xxx.lanl.gov/abs/1002.1011}{{\tt
  arXiv:1002.1011}}].

\bibitem{Alonso:2012px}
R.~Alonso, M.~B. Gavela, L.~Merlo, S.~Rigolin, and J.~Yepes,  Phys. Lett. {\bf
  B722} (2013) 330--335, [\href{http://xxx.lanl.gov/abs/1212.3305}{{\tt
  arXiv:1212.3305}}]. [Erratum: Phys. Lett.B726,926(2013)].

\bibitem{Alonso:2012pz}
R.~Alonso, M.~B. Gavela, L.~Merlo, S.~Rigolin, and J.~Yepes,  Phys. Rev. {\bf
  D87} (2013), no.~5 055019, [\href{http://xxx.lanl.gov/abs/1212.3307}{{\tt
  arXiv:1212.3307}}].

\bibitem{Buchalla:2013rka}
G.~Buchalla, O.~Cata and C.~Krause, Nucl.\ Phys.\ B {\bf 880} (2014) 552 
Erratum: [Nucl.\ Phys.\ B {\bf 913} (2016) 475], 
[\href{http://xxx.lanl.gov/abs/1307.5017}{{\tt arXiv:1307.5017}}].

\bibitem{Gavela:2014vra}
M.~B. Gavela, J.~Gonzalez-Fraile, M.~C. Gonzalez-Garcia, L.~Merlo, S.~Rigolin,
  and J.~Yepes,  JHEP {\bf 10} (2014) 044,
  [\href{http://xxx.lanl.gov/abs/1406.6367}{{\tt arXiv:1406.6367}}].

\bibitem{Alonso:2012jc}
R.~Alonso, M.~B. Gavela, L.~Merlo, S.~Rigolin, and J.~Yepes,  JHEP {\bf 06}
  (2012) 076, [\href{http://xxx.lanl.gov/abs/1201.1511}{{\tt
  arXiv:1201.1511}}].

\bibitem{Brivio:2013pma}
I.~Brivio, T.~Corbett, O.~J.~P. Eboli, M.~B. Gavela, J.~Gonzalez-Fraile, M.~C.
  Gonzalez-Garcia, L.~Merlo, and S.~Rigolin,  JHEP {\bf 03} (2014) 024,
  [\href{http://xxx.lanl.gov/abs/1311.1823}{{\tt arXiv:1311.1823}}].

\bibitem{Brivio:2014pfa}
I.~Brivio, O.~J.~P. Eboli, M.~B. Gavela, M.~C. Gonzalez-Garcia, L.~Merlo, and
  S.~Rigolin,  JHEP {\bf 12} (2014) 004,
  [\href{http://xxx.lanl.gov/abs/1405.5412}{{\tt arXiv:1405.5412}}].

\bibitem{Brivio:2015kia}
I.~Brivio, M.~B. Gavela, L.~Merlo, K.~Mimasu, J.~M. No, R.~del Rey, and
  V.~Sanz,  JHEP {\bf 04} (2016) 141,
  [\href{http://xxx.lanl.gov/abs/1511.01099}{{\tt arXiv:1511.01099}}].

\bibitem{Brivio:2016fzo}
I.~Brivio, J.~Gonzalez-Fraile, M.~C. Gonzalez-Garcia, and L.~Merlo,  Eur. Phys.
  J. {\bf C76} (2016), no.~7 416,
  [\href{http://xxx.lanl.gov/abs/1604.06801}{{\tt arXiv:1604.06801}}].

\bibitem{Merlo:2016prs}
L.~Merlo, S.~Saa, and M.~S. Barbero,
  \href{http://xxx.lanl.gov/abs/1612.04832}{{\tt arXiv:1612.04832}}.

\bibitem{Brivio:2017ije}
I.~Brivio, M.~B. Gavela, L.~Merlo, K.~Mimasu, J.~M. No, R.~del Rey, and
  V.~Sanz,  \href{http://xxx.lanl.gov/abs/1701.05379}{{\tt arXiv:1701.05379}}.
  
\bibitem{Gavela:2014uta}
M.~B. Gavela, K.~Kanshin, P.~A.~N. Machado, and S.~Saa,  JHEP {\bf 03} (2015) 043, 
[\href{http://xxx.lanl.gov/abs/1409.1571}{{\tt arXiv:1409.1571}}].

\bibitem{Manohar:1983md}
A.~Manohar and H.~Georgi,  Nucl. Phys. {\bf B234} (1984) 189.

\bibitem{Cohen:1997rt}
A.~G. Cohen, D.~B. Kaplan, and A.~E. Nelson,  Phys. Lett. {\bf B412} (1997)
  301--308, [\href{http://xxx.lanl.gov/abs/hep-ph/9706275}{{\tt
  hep-ph/9706275}}].

\bibitem{Gavela:2016bzc}
B.~M. Gavela, E.~E. Jenkins, A.~V. Manohar, and L.~Merlo,  Eur. Phys. J. {\bf
  C76} (2016), no.~9 485, [\href{http://xxx.lanl.gov/abs/1601.07551}{{\tt
  arXiv:1601.07551}}].

\bibitem{Eboli:2016kko}
O.~J.~P. Eboli and M.~C. Gonzalez-Garcia,  Phys. Rev. {\bf D93} (2016), no.~9
  093013, [\href{http://xxx.lanl.gov/abs/1604.03555}{{\tt arXiv:1604.03555}}].

\bibitem{Kaplan:1983fs}
D.~B. Kaplan and H.~Georgi,  Phys. Lett. {\bf B136} (1984) 183.

\bibitem{Kaplan:1983sm}
D.~B. Kaplan, H.~Georgi, and S.~Dimopoulos,  Phys. Lett. {\bf B136} (1984) 187.

\bibitem{Banks:1984gj}
T.~Banks,  Nucl. Phys. {\bf B243} (1984) 125.

\bibitem{Agashe:2004rs}
K.~Agashe, R.~Contino, and A.~Pomarol,  Nucl. Phys. {\bf B719} (2005) 165--187,
  [\href{http://xxx.lanl.gov/abs/hep-ph/0412089}{{\tt hep-ph/0412089}}].

\bibitem{Gripaios:2009pe}
B.~Gripaios, A.~Pomarol, F.~Riva, and J.~Serra,  JHEP {\bf 04} (2009) 070,
  [\href{http://xxx.lanl.gov/abs/0902.1483}{{\tt arXiv:0902.1483}}].

\bibitem{Feruglio:2016zvt}
F.~Feruglio, B.~Gavela, K.~Kanshin, P.~A.~N. Machado, S.~Rigolin, and S.~Saa,
  \href{http://xxx.lanl.gov/abs/1603.05668}{{\tt arXiv:1603.05668}}.

\bibitem{Gavela:2016vte}
M.~B. Gavela, K.~Kanshin, P.~A.~N. Machado, and S.~Saa,
  \href{http://xxx.lanl.gov/abs/1610.08083}{{\tt arXiv:1610.08083}}.

\bibitem{Halyo:1991pc}
E.~Halyo,  Mod. Phys. Lett. {\bf A8} (1993) 275--284.

\bibitem{Goldberger:2008zz}
W.~D. Goldberger, B.~Grinstein, and W.~Skiba,  Phys. Rev. Lett. {\bf 100}
  (2008) 111802, [\href{http://xxx.lanl.gov/abs/0708.1463}{{\tt
  arXiv:0708.1463}}].

\bibitem{Vecchi:2010gj}
L.~Vecchi,  Phys. Rev. {\bf D82} (2010) 076009,
  [\href{http://xxx.lanl.gov/abs/1002.1721}{{\tt arXiv:1002.1721}}].

\bibitem{Matsuzaki:2012mk}
S.~Matsuzaki and K.~Yamawaki,  Phys. Lett. {\bf B719} (2013) 378--382,
  [\href{http://xxx.lanl.gov/abs/1207.5911}{{\tt arXiv:1207.5911}}].

\bibitem{Chacko:2012sy}
Z.~Chacko and R.~K. Mishra,  Phys. Rev. {\bf D87} (2013), no.~11 115006,
  [\href{http://xxx.lanl.gov/abs/1209.3022}{{\tt arXiv:1209.3022}}].

\bibitem{Chacko:2012vm}
Z.~Chacko, R.~Franceschini, and R.~K. Mishra,  JHEP {\bf 04} (2013) 015,
  [\href{http://xxx.lanl.gov/abs/1209.3259}{{\tt arXiv:1209.3259}}].

\bibitem{Bellazzini:2012vz}
B.~Bellazzini, C.~Csaki, J.~Hubisz, J.~Serra, and J.~Terning,  Eur. Phys. J.
  {\bf C73} (2013), no.~2 2333, [\href{http://xxx.lanl.gov/abs/1209.3299}{{\tt
  arXiv:1209.3299}}].

\bibitem{Alonso:2014wta}
R.~Alonso, I.~Brivio, B.~Gavela, L.~Merlo, and S.~Rigolin,  JHEP {\bf 12}
  (2014) 034, [\href{http://xxx.lanl.gov/abs/1409.1589}{{\tt
  arXiv:1409.1589}}].

\bibitem{Hierro:2015nna}
I.~M. Hierro, L.~Merlo, and S.~Rigolin,  JHEP {\bf 04} (2016) 016,
  [\href{http://xxx.lanl.gov/abs/1510.07899}{{\tt arXiv:1510.07899}}].

\bibitem{Fodor:2012ty}
Z.~Fodor, K.~Holland, J.~Kuti, D.~Nogradi, C.~Schroeder, and C.~H. Wong,  Phys.
  Lett. {\bf B718} (2012) 657--666,
  [\href{http://xxx.lanl.gov/abs/1209.0391}{{\tt arXiv:1209.0391}}].

\bibitem{Aoki:2014oha}
{\bf LatKMI} Collaboration, Y.~Aoki {\em et.~al.},  Phys. Rev. {\bf D89} (2014)
  111502, [\href{http://xxx.lanl.gov/abs/1403.5000}{{\tt arXiv:1403.5000}}].

\bibitem{Appelquist:2016viq}
T.~Appelquist {\em et.~al.},  Phys. Rev. {\bf D93} (2016), no.~11 114514,
  [\href{http://xxx.lanl.gov/abs/1601.04027}{{\tt arXiv:1601.04027}}].

\bibitem{Fodor:2016pls}
Z.~Fodor, K.~Holland, J.~Kuti, S.~Mondal, D.~Nogradi, and C.~H. Wong,  PoS {\bf
  LATTICE2015} (2016) 219, [\href{http://xxx.lanl.gov/abs/1605.08750}{{\tt
  arXiv:1605.08750}}].

\bibitem{Aoki:2016wnc}
{\bf LatKMI} Collaboration, Y.~Aoki {\em et.~al.},
  \href{http://xxx.lanl.gov/abs/1610.07011}{{\tt arXiv:1610.07011}}.

\bibitem{Appelquist:2017wcg}
T.~Appelquist, J.~Ingoldby, and M.~Piai,
  \href{http://xxx.lanl.gov/abs/1702.04410}{{\tt arXiv:1702.04410}}.

\bibitem{Hellerman:2015nra}
S.~Hellerman, D.~Orlando, S.~Reffert, and M.~Watanabe,  JHEP {\bf 12} (2015)
  071, [\href{http://xxx.lanl.gov/abs/1505.01537}{{\tt arXiv:1505.01537}}].

\bibitem{Monin:2016jmo}
A.~Monin, D.~Pirtskhalava, R.~Rattazzi, and F.~K. Seibold,
  \href{http://xxx.lanl.gov/abs/1611.02912}{{\tt arXiv:1611.02912}}.

\bibitem{Grinstein:2007iv}
B.~Grinstein and M.~Trott,  Phys. Rev. {\bf D76} (2007) 073002,
  [\href{http://xxx.lanl.gov/abs/0704.1505}{{\tt arXiv:0704.1505}}].

\bibitem{Alonso:2015fsp}
R.~Alonso, E.~E. Jenkins, and A.~V. Manohar,  Phys. Lett. {\bf B754} (2016)
  335--342, [\href{http://xxx.lanl.gov/abs/1511.00724}{{\tt
  arXiv:1511.00724}}].

\bibitem{Alonso:2016btr}
R.~Alonso, E.~E. Jenkins, and A.~V. Manohar,  Phys. Lett. {\bf B756} (2016)
  358--364, [\href{http://xxx.lanl.gov/abs/1602.00706}{{\tt
  arXiv:1602.00706}}].

\bibitem{Alonso:2016oah}
R.~Alonso, E.~E. Jenkins, and A.~V. Manohar,  JHEP {\bf 08} (2016) 101,
  [\href{http://xxx.lanl.gov/abs/1605.03602}{{\tt arXiv:1605.03602}}].

\bibitem{Hirn:2005fr}
J.~Hirn and J.~Stern,  Phys. Rev. {\bf D73} (2006) 056001,
  [\href{http://xxx.lanl.gov/abs/hep-ph/0504277}{{\tt hep-ph/0504277}}].

\bibitem{Bellazzini:2014yua}
B.~Bellazzini, C.~Csáki, and J.~Serra,  Eur. Phys. J. {\bf C74} (2014), no.~5
  2766, [\href{http://xxx.lanl.gov/abs/1401.2457}{{\tt arXiv:1401.2457}}].

\bibitem{Panico:2015jxa}
G.~Panico and A.~Wulzer,  Lect. Notes Phys. {\bf 913} (2016) pp.1--316,
  [\href{http://xxx.lanl.gov/abs/1506.01961}{{\tt arXiv:1506.01961}}].

\bibitem{Holdom:1984sk}
B.~Holdom,  Phys. Lett. {\bf B150} (1985) 301--305.

\bibitem{Yamawaki:1985zg}
K.~Yamawaki, M.~Bando, and K.-i. Matumoto,  Phys. Rev. Lett. {\bf 56} (1986)
  1335.

\bibitem{Appelquist:1986an}
T.~W. Appelquist, D.~Karabali, and L.~C.~R. Wijewardhana,  Phys. Rev. Lett.
  {\bf 57} (1986) 957.

\bibitem{Randall:1999ee}
L.~Randall and R.~Sundrum,  Phys. Rev. Lett. {\bf 83} (1999) 3370--3373,
  [\href{http://xxx.lanl.gov/abs/hep-ph/9905221}{{\tt hep-ph/9905221}}].

\bibitem{Contino:2003ve}
R.~Contino, Y.~Nomura, and A.~Pomarol,  Nucl. Phys. {\bf B671} (2003) 148--174,
  [\href{http://xxx.lanl.gov/abs/hep-ph/0306259}{{\tt hep-ph/0306259}}].

\bibitem{Gherghetta:2003he}
T.~Gherghetta and A.~Pomarol,  Phys. Rev. {\bf D67} (2003) 085018,
  [\href{http://xxx.lanl.gov/abs/hep-ph/0302001}{{\tt hep-ph/0302001}}].

\bibitem{Agashe:2003zs}
K.~Agashe, A.~Delgado, M.~J. May, and R.~Sundrum,  JHEP {\bf 08} (2003) 050,
  [\href{http://xxx.lanl.gov/abs/hep-ph/0308036}{{\tt hep-ph/0308036}}].
 
\bibitem{Chacko:2013dra}
Z.~Chacko, R.~K.~Mishra and D.~Stolarski,  JHEP {\bf 1309} (2013) 121,
  [\href{http://xxx.lanl.gov/abs/1304.1795}{{\tt arXiv:1304.1795}}]. 
  
\bibitem{Bellazzini:2013fga}
B.~Bellazzini, C.~Csaki, J.~Hubisz, J.~Serra and J.~Terning,  Eur.\ Phys.\ J.\ C {\bf 74} (2014) 2790,
  [\href{http://xxx.lanl.gov/abs/1305.3919}{{\tt arXiv:1305.3919}}]. 

\bibitem{Coradeschi:2013gda}
F.~Coradeschi, P.~Lodone, D.~Pappadopulo, R.~Rattazzi and L.~Vitale,  JHEP {\bf 1311} (2013) 057,
  [\href{http://xxx.lanl.gov/abs/1306.4601}{{\tt arXiv:1306.4601}}]. 

\bibitem{Megias:2014iwa}
E.~Megias and O.~Pujolas,  JHEP {\bf 1408} (2014) 081
  [\href{http://xxx.lanl.gov/abs/1401.4998}{{\tt arXiv:1401.4998}}].   

\bibitem{ColemanBook}
S.~Coleman, ``Dilatations'', in Aspects of Symmetry: Selected Erice Lectures (pp. 67-98). Cambridge: Cambridge University Press. 1985.

\bibitem{D'Ambrosio:2002ex}
G.~D'Ambrosio, G.~F. Giudice, G.~Isidori, and A.~Strumia,  Nucl. Phys. {\bf
  B645} (2002) 155--187, [\href{http://xxx.lanl.gov/abs/hep-ph/0207036}{{\tt
  hep-ph/0207036}}].

\bibitem{Cirigliano:2005ck}
V.~Cirigliano, B.~Grinstein, G.~Isidori, and M.~B. Wise,  Nucl. Phys. {\bf
  B728} (2005) 121--134, [\href{http://xxx.lanl.gov/abs/hep-ph/0507001}{{\tt
  hep-ph/0507001}}].

\bibitem{Davidson:2006bd}
S.~Davidson and F.~Palorini,  Phys. Lett. {\bf B642} (2006) 72--80,
  [\href{http://xxx.lanl.gov/abs/hep-ph/0607329}{{\tt hep-ph/0607329}}].

\bibitem{Grinstein:2010ve}
B.~Grinstein, M.~Redi, and G.~Villadoro,  JHEP {\bf 11} (2010) 067,
  [\href{http://xxx.lanl.gov/abs/1009.2049}{{\tt arXiv:1009.2049}}].

\bibitem{Feldmann:2010yp}
T.~Feldmann,  JHEP {\bf 04} (2011) 043,
  [\href{http://xxx.lanl.gov/abs/1010.2116}{{\tt arXiv:1010.2116}}].

\bibitem{Alonso:2011yg}
R.~Alonso, M.~B. Gavela, L.~Merlo, and S.~Rigolin,  JHEP {\bf 07} (2011) 012,
  [\href{http://xxx.lanl.gov/abs/1103.2915}{{\tt arXiv:1103.2915}}].

\bibitem{Guadagnoli:2011id}
D.~Guadagnoli, R.~N. Mohapatra, and I.~Sung,  JHEP {\bf 04} (2011) 093,
  [\href{http://xxx.lanl.gov/abs/1103.4170}{{\tt arXiv:1103.4170}}].

\bibitem{Alonso:2011jd}
R.~Alonso, G.~Isidori, L.~Merlo, L.~A. Munoz, and E.~Nardi,  JHEP {\bf 06}
  (2011) 037, [\href{http://xxx.lanl.gov/abs/1103.5461}{{\tt
  arXiv:1103.5461}}].

\bibitem{Buras:2011zb}
A.~J. Buras, L.~Merlo, and E.~Stamou,  JHEP {\bf 08} (2011) 124,
  [\href{http://xxx.lanl.gov/abs/1105.5146}{{\tt arXiv:1105.5146}}].

\bibitem{Buras:2011wi}
A.~J. Buras, M.~V. Carlucci, L.~Merlo, and E.~Stamou,  JHEP {\bf 03} (2012)
  088, [\href{http://xxx.lanl.gov/abs/1112.4477}{{\tt arXiv:1112.4477}}].

\bibitem{Alonso:2012fy}
R.~Alonso, M.~B. Gavela, D.~Hernandez, and L.~Merlo,  Phys. Lett. {\bf B715}
  (2012) 194--198, [\href{http://xxx.lanl.gov/abs/1206.3167}{{\tt
  arXiv:1206.3167}}].

\bibitem{Lopez-Honorez:2013wla}
L.~Lopez-Honorez and L.~Merlo,  Phys. Lett. {\bf B722} (2013) 135--143,
  [\href{http://xxx.lanl.gov/abs/1303.1087}{{\tt arXiv:1303.1087}}].

\bibitem{Alonso:2013mca}
R.~Alonso, M.~B. Gavela, D.~Hernandez, L.~Merlo, and S.~Rigolin,  JHEP {\bf
  08} (2013) 069, [\href{http://xxx.lanl.gov/abs/1306.5922}{{\tt
  arXiv:1306.5922}}].

\bibitem{Alonso:2016onw}
R.~Alonso, E.~F. Mart{\'\i ne}z, M.~B. Gavela, B.~Grinstein, L.~Merlo, and
  P.~Quilez,  JHEP {\bf 12} (2016) 119,
  [\href{http://xxx.lanl.gov/abs/1609.05902}{{\tt arXiv:1609.05902}}].
  
\bibitem{Dinh:2017smk}
D.~Dinh, L.~Merlo, S.~T.~Petcov and Vega-\'Alvarez,
\href{http://xxx.lanl.gov/abs/1705.09284}{{\tt arXiv:1705.09284}}

\bibitem{Froggatt:1978nt}
C.~D. Froggatt and H.~B. Nielsen,  Nucl. Phys. {\bf B147} (1979) 277.

\bibitem{Ma:2004zv}
E.~Ma,  Phys. Rev. {\bf D70} (2004) 031901,
  [\href{http://xxx.lanl.gov/abs/hep-ph/0404199}{{\tt hep-ph/0404199}}].

\bibitem{Altarelli:2005yp}
G.~Altarelli and F.~Feruglio,  Nucl. Phys. {\bf B720} (2005) 64--88,
  [\href{http://xxx.lanl.gov/abs/hep-ph/0504165}{{\tt hep-ph/0504165}}].

\bibitem{Altarelli:2005yx}
G.~Altarelli and F.~Feruglio,  Nucl. Phys. {\bf B741} (2006) 215--235,
  [\href{http://xxx.lanl.gov/abs/hep-ph/0512103}{{\tt hep-ph/0512103}}].

\bibitem{Feruglio:2007uu}
F.~Feruglio, C.~Hagedorn, Y.~Lin, and L.~Merlo,  Nucl. Phys. {\bf B775} (2007)
  120--142, [\href{http://xxx.lanl.gov/abs/hep-ph/0702194}{{\tt
  hep-ph/0702194}}]. [Erratum: Nucl. Phys.B836,127(2010)].

\bibitem{Feruglio:2008ht}
F.~Feruglio, C.~Hagedorn, Y.~Lin, and L.~Merlo,  Nucl. Phys. {\bf B809} (2009)
  218--243, [\href{http://xxx.lanl.gov/abs/0807.3160}{{\tt arXiv:0807.3160}}].

\bibitem{Bazzocchi:2009pv}
F.~Bazzocchi, L.~Merlo, and S.~Morisi,  Nucl. Phys. {\bf B816} (2009) 204--226,
  [\href{http://xxx.lanl.gov/abs/0901.2086}{{\tt arXiv:0901.2086}}].

\bibitem{Altarelli:2009gn}
G.~Altarelli, F.~Feruglio, and L.~Merlo,  JHEP {\bf 05} (2009) 020,
  [\href{http://xxx.lanl.gov/abs/0903.1940}{{\tt arXiv:0903.1940}}].

\bibitem{Altarelli:2010gt}
G.~Altarelli and F.~Feruglio,  Rev. Mod. Phys. {\bf 82} (2010) 2701--2729,
  [\href{http://xxx.lanl.gov/abs/1002.0211}{{\tt arXiv:1002.0211}}].

\bibitem{Varzielas:2010mp}
I.~de~Medeiros~Varzielas and L.~Merlo,  JHEP {\bf 02} (2011) 062,
  [\href{http://xxx.lanl.gov/abs/1011.6662}{{\tt arXiv:1011.6662}}].

\bibitem{Toorop:2010ex}
R.~de~Adelhart~Toorop, F.~Bazzocchi, L.~Merlo, and A.~Paris,  JHEP {\bf 03}
  (2011) 035, [\href{http://xxx.lanl.gov/abs/1012.1791}{{\tt
  arXiv:1012.1791}}]. [Erratum: JHEP01,098(2013)].

\bibitem{Altarelli:2012bn}
G.~Altarelli, F.~Feruglio, L.~Merlo, and E.~Stamou,  JHEP {\bf 08} (2012) 021,
  [\href{http://xxx.lanl.gov/abs/1205.4670}{{\tt arXiv:1205.4670}}].

\bibitem{Altarelli:2012ss}
G.~Altarelli, F.~Feruglio, and L.~Merlo,  Fortsch. Phys. {\bf 61} (2013)
  507--534, [\href{http://xxx.lanl.gov/abs/1205.5133}{{\tt arXiv:1205.5133}}].

\bibitem{Bazzocchi:2012st}
F.~Bazzocchi and L.~Merlo,  Fortsch. Phys. {\bf 61} (2013) 571--596,
  [\href{http://xxx.lanl.gov/abs/1205.5135}{{\tt arXiv:1205.5135}}].

\bibitem{Altarelli:2012ia}
G.~Altarelli, F.~Feruglio, I.~Masina, and L.~Merlo,  JHEP {\bf 11} (2012) 139,
  [\href{http://xxx.lanl.gov/abs/1207.0587}{{\tt arXiv:1207.0587}}].

\bibitem{Bergstrom:2014owa}
J.~Bergstrom, D.~Meloni, and L.~Merlo,  Phys. Rev. {\bf D89} (2014), no.~9
  093021, [\href{http://xxx.lanl.gov/abs/1403.4528}{{\tt arXiv:1403.4528}}].
  
\bibitem{Giudice:2007fh}
G.~F.~Giudice, C.~Grojean, A.~Pomarol and R.~Rattazzi,  JHEP {\bf 0706} (2007) 045, 
[\href{http://xxx.lanl.gov/abs/hep-ph/0703164}{{\tt arXiv:hep-ph/0703164}}]. 

\bibitem{Ballestrero:2009vw}
A.~Ballestrero, G.~Bevilacqua, D.~Buarque Franzosi and E.~Maina, JHEP {\bf 0911} (2009) 126,
[\href{http://xxx.lanl.gov/abs/hep-ph/0909.3838}{{\tt arXiv:hep-ph/0909.3838}}]. 
  
\bibitem{Longhitano:1980iz}
A.~C.~Longhitano, Phys.\ Rev.\ D {\bf 22} (1980) 1166.

\bibitem{Longhitano:1980tm}
A.~C.~Longhitano, Nucl.\ Phys.\ B {\bf 188} (1981) 118.

\bibitem{Espriu:2013fia}
D.~Espriu, F.~Mescia and B.~Yencho, Phys.\ Rev.\ D {\bf 88} (2013) 055002,
[\href{http://xxx.lanl.gov/abs/1307.2400}{{\tt 1307.2400}}].

\bibitem{Delgado:2013loa}
R.~L.~Delgado, A.~Dobado and F.~J.~Llanes-Estrada, J.\ Phys.\ G {\bf 41} (2014) 025002,
[\href{http://xxx.lanl.gov/abs/1308.1629}{{\tt 1308.1629}}].

\bibitem{Delgado:2014jda}
R.~L.~Delgado, A.~Dobado, M.~J.~Herrero and J.~J.~Sanz-Cillero, JHEP {\bf 1407} (2014) 149,
[\href{http://xxx.lanl.gov/abs/1404.2866}{{\tt 1404.2866}}].
  
\bibitem{Guo:2015isa}
F.~K.~Guo, P.~Ruiz-Femenía and J.~J.~Sanz-Cillero, Phys.\ Rev.\ D {\bf 92} (2015) 074005,
[\href{http://xxx.lanl.gov/abs/1506.04204}{{\tt arXiv:1506.04204}}].

\bibitem{Hagiwara:1993ck}
K.~Hagiwara, S.~Ishihara, R.~Szalapski, and D.~Zeppenfeld,  Phys. Rev. {\bf
  D48} (1993) 2182--2203.

\bibitem{Hagiwara:1996kf}
K.~Hagiwara, T.~Hatsukano, S.~Ishihara, and R.~Szalapski,  Nucl. Phys. {\bf
  B496} (1997) 66--102, [\href{http://xxx.lanl.gov/abs/hep-ph/9612268}{{\tt
  hep-ph/9612268}}].

\bibitem{DeRujula:1991ufe}
A.~De~Rujula, M.~B. Gavela, P.~Hernandez, and E.~Masso,  Nucl. Phys. {\bf B384} (1992) 3--58.


\end{thebibliography}

\providecommand{\href}[2]{#2}\begingroup\raggedright\endgroup

\end{document}